\newcommand{\change}[1]{#1}

\documentclass[sn-mathphys]{sn-jnl}

\newcommand{\V}[1]{\mathbf{#1}} 
\newcommand\Alfven{Alfv\'en} 

\newcommand{\ua}{\orgdiv{Lunar and Planetary Laboratory}, \orgname{University of Arizona}, \orgaddress{\city{Tucson}, \state{AZ}  \postcode{85721}, \country{USA}}}
\newcommand{\unh}{\orgdiv{Space Science Center}, \orgname{University of New Hampshire}, \orgaddress{\city{Durham}, \state{NH} \postcode{03824}, \country{USA}}}
\newcommand{\lesia}{\orgdiv{Observatoire de Paris}, \orgname{LESIA}, \orgaddress{\city{Meudon}, \postcode{92190}, \country{France}}}
\newcommand{\ias}{\orgdiv{Institute for Advanced Study}, \orgaddress{\city{Princeton}, \state{NJ}, \postcode{08540}, \country{USA}}}
\newcommand{\uchicago}{\orgdiv{Department of Astronomy \& Astrophysics}, \orgname{University of Chicago} \orgaddress{\city{Chicago}, \state{IL}, \postcode{60637}, \country{USA}}}
\newcommand{\bwxt}{\orgname{BWX Technologies, Inc.} \city{Washington DC}, \postcode{20001}, \country{USA}}
\newcommand{\gsfc}{\orgdiv{Heliophysics Science Division}, \orgname{NASA Goddard Space Flight Center}, \city{Greenbelt}, \state{MD}, \postcode{20771}, \country{USA}}
\newcommand{\imperial}{\orgdiv{Blackett Laboratory}, \orgname{Imperial College London} \city{London}, \postcode{SW7 2AZ}, \country{UK}}
\newcommand{\mssl}{\orgdiv{Mullard Space Science Laboratory}, \orgname{University College London} \city{Dorking}, \postcode{RH5\,6NT}, \country{UK}}
\newcommand{\irap}{\orgdiv{Institut de Recherche en en Astrophysique et Planetologie}, 31028 \city{Toulouse}, \country{France}}
\newcommand{\uiowa}{\orgdiv{Department of Physics and Astronomy}, \orgname{University of Iowa}, \city{Iowa City}, \state{IA} \postcode{52242}, \country{USA}}
\newcommand{\ames}{\orgdiv{Ames Research Center}, \orgname{NASA}, \city{Mountain View}, \state{CA}, \postcode{94043}, \country{USA}}
\newcommand{\lpce}{\orgdiv{Laboratoire de Physique et Chimie de l’Environnement et de l’Espace}, \orgname{CNRS and Université d'Orléans}, 45071 \city{Orl\'eans}, \country{France}}
\newcommand{\princeton}{\orgdiv{Department of Astrophysical Sciences}, \orgname{Princeton University} \city{Princeton}, \state{NJ}, \postcode{08540} \country{USA}}
\newcommand{\lab}{\orgdiv{Laboratoire d'astrophysique de Bordeaux, Univ. Bordeaux, CNRS}, \city{Pessac}, \country{France}}
\newcommand{\lpp}{\orgdiv{Laboratoire de Physique des Plasmas}, \orgname{CNRS/Sorbonne Université/Université Paris-Saclay/Observatoire de Paris/Ecole Polytechnique Institut Polytechnique de Paris}, \city{Paris}, \postcode{75005} \country{France}}
\newcommand{\ssl}{\orgdiv{Space Sciences Laboratory}, \orgname{University of California Berkeley}, \orgaddress{\city{Berkeley}, \state{CA}  \postcode{94720}, \country{USA}}}
\newcommand{\udel}{\orgdiv{Department of Physics \& Astronomy}, \orgname{University of Delaware}, \orgaddress{\city{Newark}, \state{DE}  \postcode{19716}, \country{USA}}}
\newcommand{\cambridge}{\orgdiv{Institute of Astronomy}, \orgname{University of Cambridge}, \orgaddress{\city{Cambridge}, \postcode{CB3 OHA}, \country{UK}}}
\newcommand{\oeaw}{Space Research Institute, Austrian Academy of Sciences, Graz, Austria}
\newcommand{\oxford}{\orgdiv{Rudolf Peierls Centre for Theoretical Physics}, \orgname{University of Oxford}, \orgaddress{\city{Oxford}, \postcode{OX1 3PU}, \country{UK}}}
\newcommand{\lanl}{\orgdiv{Los Alamos National Laboratory}, \orgaddress{\city{Las Alamos}, \state{NM} \postcode{87545}, \country{USA}}}
\newcommand{\cfa}{\orgdiv{Center for Astrophysics}, \orgname{Harvard \& Smithsonian} \orgaddress{\city{Cambridge}, \state{MA} \postcode{02138}, \country{USA}}}
\newcommand{\uah}{\orgdiv{Department of Space Science}, \orgname{University of Alabama in Huntsville} \orgaddress{\city{Huntsville}, \state{AL} \postcode{35899}, \country{USA}}}
\newcommand{\uwm}{\orgdiv{Department of Astronomy}, \orgname{University of Wisconsin-Madison} \orgaddress{\city{Madison}, \state{WI} \postcode{53706}, \country{USA}}}

\usepackage{amssymb}
\usepackage{amsmath}
\usepackage{hyperref}
\usepackage{natbib}
\usepackage{color}
\usepackage{graphicx}
\usepackage{subfig}

\begin{document}

\title[HelioSwarm]{HelioSwarm: A Multipoint, Multiscale Mission to Characterize Turbulence}

\author*[1]{\fnm{Kristopher} G. \sur{Klein}}\email{kgklein@arizona.edu} 
\author*[2]{\fnm{Harlan} \sur{Spence}}\email{harlan.spence@unh.edu} 
\author[3]{\fnm{Olga} \sur{Alexandrova}}
\author[2]{\fnm{Matthew} \sur{Argall}}
\author[4]{\fnm{Lev} \sur{Arzamasskiy}}
\author[12]{\fnm{Jay} \sur{Bookbinder}}
\author[1]{\fnm{Theodore} \sur{Broeren}}
\author[5]{\fnm{Damiano} \sur{Caprioli}}
\author[6]{\fnm{Anthony} \sur{Case}}
\author[2]{\fnm{Benjamin} \sur{Chandran}}
\author[7]{\fnm{Li-Jen} \sur{Chen}}
\author[2]{\fnm{Ivan} \sur{Dors}}
\author[8]{\fnm{Jonathan} \sur{Eastwood}}
\author[9]{\fnm{Colin} \sur{Forsyth}}
\author[2]{\fnm{Antoinette} \sur{Galvin}}
\author[10]{\fnm{Vincent} \sur{Genot}}
\author[11]{\fnm{Jasper} \sur{Halekas}$^{\textrm{11}}$}
\author[12]{\fnm{Michael} \sur{Hesse}}
\author[12]{\fnm{Butler} \sur{Hine}}
\author[8]{\fnm{Tim} \sur{Horbury}}
\author[7]{\fnm{Lan} \sur{Jian}}
\author[6]{\fnm{Justin} \sur{Kasper}}
\author[13]{\fnm{Matthieu} \sur{Kretzschmar}}
\author[14]{\fnm{Matthew} \sur{Kunz}}
\author[10,15]{\fnm{Benoit} \sur{Lavraud}}
\author[16]{\fnm{Olivier} \sur{Le Contel}}
\author[17]{\fnm{Alfred} \sur{Mallet}}
\author[18]{\fnm{Bennett} \sur{Maruca}}
\author[18]{\fnm{William} \sur{Matthaeus}}
\author[2]{\fnm{Jonathan} \sur{Niehof}}
\author[8]{\fnm{Helen} \sur{O'Brian}}
\author[9]{\fnm{Christopher} \sur{Owen}}
\author[16]{\fnm{Alessandro} \sur{Retino}}
\author[19]{\fnm{Christopher} \sur{Reynolds}}
\author[20]{\fnm{Owen} \sur{Roberts}}
\author[21]{\fnm{Alexander} \sur{Schekochihin}}
\author[22]{\fnm{Ruth} \sur{Skoug}$^{\textrm{22}}$}
\author[2]{\fnm{Charles} \sur{Smith}}
\author[2]{\fnm{Sonya} \sur{Smith}}
\author[22]{\fnm{John} \sur{Steinberg}$^{\textrm{22}}$}
\author[23]{\fnm{Michael} \sur{Stevens}}
\author[7]{\fnm{Adam} \sur{Szabo}}
\author[14]{\fnm{Jason} \sur{TenBarge}}
\author[4]{\fnm{Roy} \sur{Torbert}}
\author[4]{\fnm{Bernard} \sur{Vasquez}}
\author[9]{\fnm{Daniel} \sur{Verscharen}}
\author[17]{\fnm{Phyllis} \sur{Whittlesey}}
\author[12]{\fnm{Brittany} \sur{Wickizer}}
\author[24]{\fnm{Gary} \sur{Zank}}
\author[25]{\fnm{Ellen} \sur{Zweibel}}

\affil[1]\ua
\affil[2]\unh
\affil[3]\lesia
\affil[4]\ias
\affil[5]\uchicago
\affil[6]\bwxt
\affil[7]\gsfc
\affil[8]\imperial
\affil[9]\mssl
\affil[10]\irap
\affil[11]\uiowa
\affil[12]\ames
\affil[13]\lpce
\affil[14]\princeton
\affil[15]\lab
\affil[16]\lpp
\affil[17]\ssl
\affil[18]\udel
\affil[19]\cambridge
\affil[20]\oeaw
\affil[21]\oxford
\affil[22]\lanl
\affil[23]\cfa
\affil[24]\uah
\affil[25]\uwm


\abstract{
HelioSwarm (HS) is a NASA Medium-Class Explorer mission of the Heliophysics Division designed to explore the dynamic three-dimensional mechanisms controlling the physics of plasma turbulence, a ubiquitous process occurring in the heliosphere and in plasmas throughout the universe. 
This will be accomplished by making simultaneous measurements at nine spacecraft with separations spanning magnetohydrodynamic and sub-ion spatial scales in a variety of near-Earth plasmas.
In this paper, we describe the scientific background for the HS investigation, the mission goals and objectives, the observatory reference trajectory and instrumentation implementation before the start of Phase B.
Through multipoint, multiscale measurements, HS promises to reveal how energy is transferred across scales and boundaries in plasmas throughout the universe.
}

\maketitle

\section{Introduction}

Turbulence is multiscale disorder.
It is the process by which energy that has been injected into a system is transported between fluctuating magnetic fields and plasma motion with larger and smaller spatial scales.
Once this cascade of energy reaches sufficiently small scales, dissipation mechanisms can act efficiently to remove energy from the fluctuations, leading to heating of the constituent particles.
Observations from single spacecraft provide only information along a single path through a turbulent system; leveraging such measurements to understand turbulence relies on assumptions about the underlying spatial and temporal structure. 
Clusters of four spacecraft provide more information about spatial structure, but are sensitive to only a single scale for a given configuration.
Understanding fundamental processes such as turbulence requires characterizing the underlying fluctuations and their dynamic evolution across many characteristic scales simultaneously. 
HelioSwarm (HS) is a Heliophysics Division NASA Medium-Class Explorer mission designed to make such multiscale observations.

\begin{figure}
    \centering
        \includegraphics{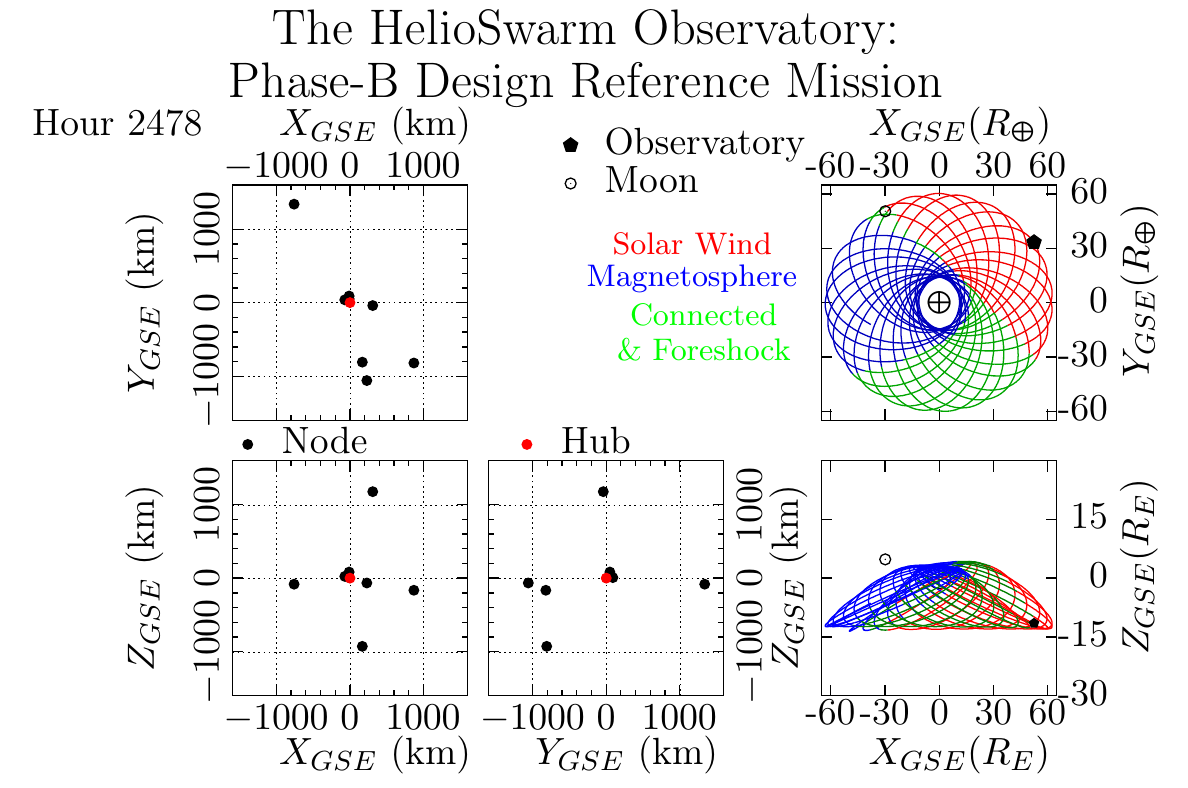}
    \caption{HelioSwarm Observatory Configuration drawn from the Phase B Design Reference Mission (DRM) trajectory. \change{In the right row, the 
    Observatory location (black pentagon) and orbits relative to Earth over the 12-month science phase are shown in both the X-Y and X-Z GSE planes.} 
    Lunar position is indicated by an open circle for reference, with different regions of near-Earth plasmas indicated with color. \change{The connected foreshock is here defined based on a typical Parker spiral orientation.}
    The remaining panels characterize two-dimensional projections of the relative configurations of the eight Nodes (black) with respect to the central Hub (red) in Geocentric Solar Ecliptic (GSE) co\"ordinates. 
    A video of the HS DRM Configuration throughout the Science Phase is available in Online Resource 1.}
    \label{fig:quad}
\end{figure}

HS, currently in Phase B-prep, is the first mission that will make the required measurements to transform the current understanding of space plasma turbulence, using a first-ever swarm of nine spacecraft (SC), composed of one \textit{Hub} and eight \textit{Nodes}. 
The nine spacecraft, comprising the \textit{HelioSwarm Observatory}, co-orbit in a lunar resonant Earth orbit, with a 2-week period, a mean-apogee radius of $\sim 60 R_E$ and a mean-perigee radius of $\sim 11 R_E$, where $R_E =6.371 \times 10^6$ m is the Earth's radius. 
This orbit, illustrated in Figure~\ref{fig:quad}, enables measurements of a variety of near-Earth plasma environments, including the pristine solar wind (SW), the magnetically connected foreshock, the magnetosheath, and the magnetosphere.
Carefully designed trajectories produce separations between the spacecraft spanning magnetohydrodynamic (MHD) and sub-ion (e.g., ion gyroradius) spatial scales, allowing us to address a broad set of questions about the three-dimensional dynamics of magnetized turbulence.
{Answering these open questions was identified as a science priority in the 2013 Heliophysics Decadal Survey\cite{heliodecadal:2013} and is deeply rooted in decades-earlier recommendations by the space science community in the 1980 report by the Plasma Turbulence Explorer Study Group\cite{Montgomery:1980}}.
As the first multipoint, multiscale mission, HS gives an unprecedented view into the nature of space plasma turbulence.


In Sec.~\ref{sec:background}, we describe the scientific motivation for HS and in Sec.~\ref{sec:goals} we enumerate the specific mission goals and objectives.
Secs.~\ref{sec:requirements} and \ref{sec:instruments} describe the requirements on what the missions will measure and the observatory trajectories and instrumentation.
Sec.~\ref{sec:analysis.methods} illustrates the application of analysis methods to synthetic data modeling future observations drawn from numerical simulations of turbulence.
Conclusions and ongoing work towards the scheduled launch date at the end of this decade are discussed in Sec.~\ref{sec:conclusions}.

\section{Scientific Background and Motivation}
\label{sec:background}

Turbulent systems consist of fluctuations spanning a wide range of spatial and temporal scales.
Fluctuations interact nonlinearly, typically with a net transfer of energy from larger to smaller spatial scales. 
This process, the energy cascade, couples the \textit{injection} range of scales, through a lossless \textit{inertial} range, into a \textit{dissipation} range where heating occurs. 
Turbulence in plasmas is significantly more complex than in hydrodynamics: plasma motion couples to dynamically significant electromagnetic fields, the system possesses many characteristic spatial and temporal scales, supports many different waves and fluctuations, and in weakly collisional systems many mechanisms other than viscosity can act to dissipate the cascade.
Additional details on the current state of plasma turbulence research can be found in recent reviews, e.g. \cite{Bruno:2013,Kiyani:2015,Verscharen:2019}. 
The most energetic SW fluctuations are non-compressive \cite{Alexandrova:2008} with properties resembling Alfv\'en waves \cite{Belcher:1971,Matthaeus:1999,Howes:2015}. 
Different types of fluctuations nonlinearly interact in different ways \cite{Zank:1996,Matthaeus:1999,Schekochihin:2009,Kunz:2015,Kunz:2018} resulting in dramatically different outcomes. 
The ubiquity of turbulence in space and astrophysical plasmas makes it a leading candidate for the process governing the thermodynamics of a wide range of systems.
For instance, turbulence is conjectured to enable angular momentum transport in accretion disks \citep{Balbus:1998}, amplify galactic magnetic fields \citep{Kulsrud:2008}, affect transport processes \cite{Kunz:2022} and establish high temperatures \citep{Zhuravleva:2018} in the intracluster medium of galaxy clusters, determine the dispersal and mixing of elements in the Interstellar Medium (ISM) \citep{Scalo:2004}, and play a key role in star formation \citep{McKee:2007}.

SW turbulence at the injection scales is driven by large-scale structures\cite{Smith:1995,Matthaeus:2007}. 
Recent observations from Parker Solar Probe (PSP) indicate that this regime is formed due to SW processing in the near-Sun environment \cite{Huang:2023,Davis:2023}.
 Measurements of the scale-to-scale rate of energy transfer, the cascade rate, near the end of the injection range generally agree with rates near the start of the inertial range, as has been explicitly demonstrated with Magnetospheric Multiscale (MMS)  observations \cite{Bandyopadhyay:2018}.

\begin{figure}
\begin{center}
\subfloat{\includegraphics[height=20ex]{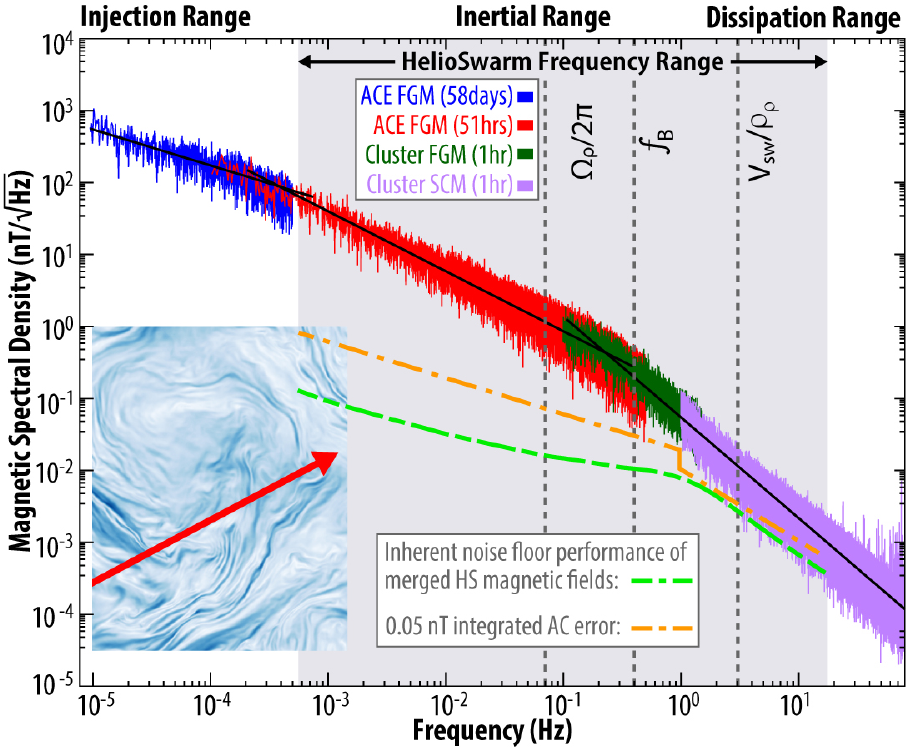}}
\subfloat{\includegraphics[height=25ex]{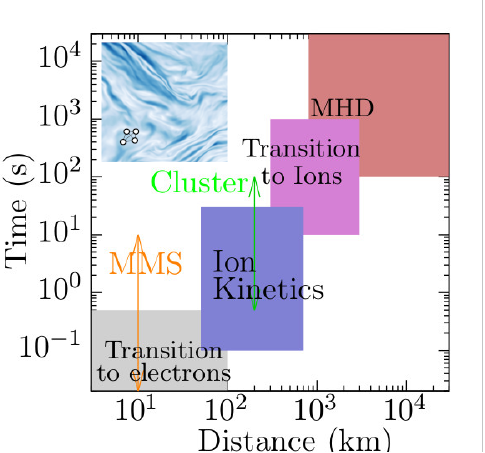}}
\subfloat{\includegraphics[height=25ex]{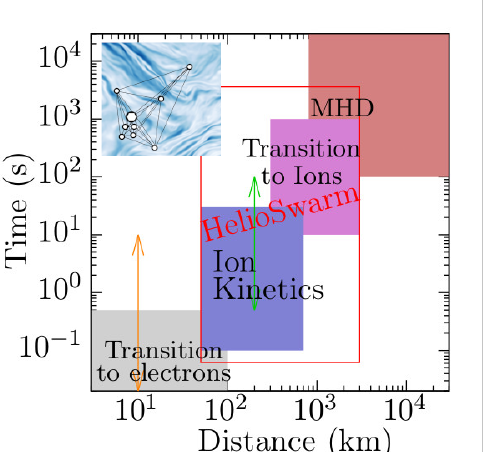}}\\
    \caption{Single spacecraft missions only provide statistical properties of SW turbulence averaged over both long times and different kinds of turbulence. 
    This approach relies on Taylor’s hypothesis to map observed time series to advected structures, measuring only a single 1D slice of the turbulence (red line in inset) and thus only provides a crude measure of turbulent properties. 
    Previous multipoint missions, e.g. MMS and Cluster, are only able to characterize spatial structure at a single-scale. 
    The HS observatory will encompass MHD and ion scales scales simultaneously, enabling the characterization of multiscale structure and dynamics of turbulence in near-Earth plasmas. Adapted from \cite{Verscharen:2019} and \cite{Arzamasskiy:2019}.} 
    \label{fig:spectra}
    \end{center}
\end{figure}

Inertial-range observations \cite{Matthaeus:1982c} exhibit scale-invariant energy transfer consistent with Kolmogorov theory \cite{Kolmogorov:1941,Kolmogorov:1962}: turbulent structures splitting into ever-smaller fluctuations while conserving energy. 
The inertial range plasma behaves like a MHD fluid \cite{Davidson:2001}, with 
MHD turbulence theory describing relevant phenomena in space physics and astrophysics \cite{Moffatt:2019,Parker:1979,Kulsrud:2005} and predicting some SW features \cite{Matthaeus:2011,Horbury:2012,Verscharen:2019}. 
For example, Figure~\ref{fig:spectra} shows a composite interplanetary magnetic field (IMF) power spectrum from three magnetometers at 1 AU measured over different time intervals from tens of days to an hour as the SW rapidly sweeps past the spacecraft. 

MHD theory adequately describes the inertial range spectral slope, but provides no guidance in the critical higher-frequency transition connecting inertial and dissipation ranges, which begins near the proton gyrofrequency $f_p \equiv \Omega_p/2\pi$.
At observed frequencies of $f_{\textrm{break}} \sim 0.33$ Hz in the SW at 1 AU---approximately equivalent to advected length scales of $L_{\textrm{break}} = v_{\textrm{SW}} /f_{\textrm{break}}\sim 1200$ km, the inertial range scale-invariance ends. 
This break arises before sub-ion scales (e.g., ion gyroradius, $\rho_p$ ), typically at apparent frequencies of $v_{\textrm{SW}} / \rho_p \sim 3$ Hz (length scales $\sim 100$ km) \cite{Klein:2019}. 
In Figure~\ref{fig:spectra} this breakdown is seen as a change in spectral slope at the transition between inertial and dissipation ranges. 
The spectral break suggests a change in the dominant physical processes and a loss of cascaded energy. 
The energy removed from the cascade is partitioned between the ions and electrons, with its dissipation leading to the heating of these charged particles.

This process of turbulent dissipation is why SW plasma is much hotter than simple theories of adiabatic expansion would predict \cite{Marsch:2012}. 
If the SW were an adiabatically expanding ideal gas, the protons at Earth would be much cooler than observed \cite{Smith:1995} and protons at Jupiter orbit ($\sim 5$ AU) would be $8$ times cooler than at 1 AU, in contrast to Voyager observations \cite{Richardson:1995}. 
Non-adiabatic heating via turbulence dominates plasma thermodynamics throughout much of the solar system, and is a leading candidate for accelerating the SW \cite{Cranmer:2012,Verdini:2010}. 

The exact heating mechanisms leading to this heating are a matter of substantial debate.
Determining the nature of these mechanisms requires observing 3D distributions of the turbulent fluctuations.
Plasma turbulence is inherently anisotropic due to preferred directions associated with the IMF \cite{Boldyrev:2006,Schekochihin:2009}, radial expansion\cite{Woodham:2021b} and large-scale gradients \cite{Volk:1972,Grappin:1993,Greco:2012b}. 
If turbulent fluctuations vary primarily parallel to the IMF (slab-like) \cite{Ghosh:1998a}, then non-compressive, Alfvénic fluctuations would, at small scales, ultimately dissipate energy via ion-cyclotron wave (ICW) heating \cite{Kasper:2013}. 
However, for fluctuations that vary mostly perpendicular to the IMF (quasi-2D \cite{Matthaeus:1990} or “critically balanced” \cite{Mallet:2015}), ICW heating is exceedingly weak. 
In this regime, dissipation instead occurs via other mechanisms such as Landau damping \cite{TenBarge:2013a} or stochastic heating \cite{Chandran:2010a}. 
Recent work on imbalanced cascades \cite{Meyrand:2021,Squire:2022} complicates these models by providing a pathway for low-frequency, anisotropic turbulence fluctuations to develop small scale structure parallel to the magnetic field, enabling dissipation via ICWs.
If turbulent structures are highly anisotropic sheets, they may undergo magnetic reconnection, which causes heating and particle acceleration \cite{Matthaeus:1986a,Mallet:2017b}. 
To distinguish among these requires an accurate determination of the 3D power distribution.
Previous determinations using single spacecraft use long time series for sufficient statistics, 
\cite{Horbury:2012,Chen:2011a,Chen:2012b,Chen:2016}, combining together intervals of turbulence with very different properties.
The regulation of energetic particle transport, in both SW \cite{Jokipii:1972} and astrophysical plasmas \cite{Zweibel:2013}, is also sensitive to the turbulence spectrum and its anisotropies.

At a fundamental level, the nature of turbulent fluctuations in magnetized plasmas remains unknown: is it an MHD extension of hydrodynamic eddies \cite{Matthaeus:2011}, a quasi-2D system \cite{Zank:1992}, critically balanced wave-like fluctuations \cite{Schekochihin:2009,Mallet:2015}, or a dynamically evolving mixture?
The complexity of plasma turbulence precludes simple, analytic solutions. 
Numerical simulations are invaluable but limited by incomplete physics and small system size\cite{Parashar:2015}. 
Confined laboratory plasmas \cite{Brown:2015,Forest:2015,Gekelman:2016} have similarly limited scale separations and access to SW-like physical parameters.

The SW is a natural laboratory where we can finally answer these questions by concurrently observing turbulent energy transfer and ion heating over a targeted range of scales. 
However, single-spacecraft observations of SW turbulence are fundamentally limited.
Multi-spacecraft missions enabled advances by creating geometric configurations to sample single-scale plasma structure without relying on Taylor’s hypothesis (futher discussed in Section~\ref{sssec:G1O1}). 
While four- (Cluster\cite{Escoubet:2001}, MMS\cite{Burch:2016}) and five-spacecraft (Time History of Events and Macroscale Interactions during Substorms (THEMIS)\cite{Angelopoulos:2008}) missions produce configurations that allow for single-scale measurements \cite{Chen:2019:WP,Escoubet:2021}, they still cannot explore multiple scales simultaneously in three dimensions. 
Even with advanced analysis techniques, scales sampled using four spacecraft cover at most a factor of $\sim 10$, \change{as demonstrated for instance with the wave-telescope technique} \cite{Sahraoui:2010a,Sahraoui:2010b}, nowhere near the $>2$ orders of magnitude necessary to simultaneously measure across inertial and dissipation ranges. 
HS’s configurations created by 9 spacecraft provide the first simultaneous multiscale view of plasma turbulence, targeting key scales from MHD to sub-ion scales.
By measuring plasmas at multiple scales simultaneously, the HS Observatory promises transformative impacts in our understanding of turbulence, which will be a boon for heliophysics, astrophysics, and plasma physics \cite{Armstrong:1981,Elmegreen:2004,MacLow:2000}.

\section{HelioSwarm Goals and Objectives}
\label{sec:goals}

HS advances Goal 4 of the 2013 National Academy of Sciences (NAS) Heliophysics Decadal Survey \cite{heliodecadal:2013} (DS) which calls on the community to "[d]iscover and characterize fundamental processes that occur both within the heliosphere and throughout the universe.”
Magnetized plasma turbulence is the primary mechanism responsible for transforming energy injected at largest scales into small-scale motions, eventually dissipating as plasma heat.
Plasma turbulence is universal, responsible for energy transfer in such diverse systems as the solar corona, SW, pulsar wind nebulae, accretion discs, interstellar medium, planet formation regions, and laboratory fusion devices.
Only the SW is both of sufficient size for multiscale observations and accessible for \textit{in situ} measurements.
Turbulence is identified as one of eight DS Goals for SW/Magnetosphere Interactions (SWMI):
“Understand the origins and effects of turbulence and wave particle interactions.” 
Because of that importance, turbulence is also identified as a SWMI Decadal Imperative: “Implement...a multi-spacecraft mission to address cross-scale plasma physics.” 
Likewise, the NASA Heliophysics Roadmap \cite{Roadmap:2014} highlights “Understand[ing] the role of turbulence and waves in the transport of mass, momentum, and energy” as one of its key Research Focus Areas of high priority. 
Long standing heliophysics mysteries --- as how the solar coronal temperature increases by orders of magnitude and how the SW is accelerated and heated --- remain unanswered after decades of research because we lack detailed understanding of how energy in turbulent plasmas heat particles.
HS advances these NAS and NASA science priorities, and will specifically resolve six science objectives (O) associated with two overarching science goals (G).

\begin{itemize}
    \item \textbf{(G1)} Reveal the 3D spatial structure and dynamics of turbulence in a weakly collisional plasma.
    \begin{itemize}
        \item \textit{G1O1} Reveal how turbulence energy transfers in the typical SW plasma as a function of scale and time.
        \item \textit{G1O2} Reveal how the turbulent cascade of energy varies with background parameters in different SW environments.
        \item \textit{G1O3} Quantify the transfer of turbulent energy between fields, flows, and proton heat.
        \item \textit{G1O4} Identify the thermodynamic impacts of intermittent structures on protons.
    \end{itemize}
    \item \textbf{(G2)} Ascertain the mutual impacts of turbulence, variability, and boundaries near large scale structures.
    \begin{itemize}
        \item \textit{G2O1} Determine how SW turbulence affects and is affected by large-scale structures such as Coronal Mass Ejections (CMEs) and Corotating Interaction Regions (CIRs).
        \item \textit{G2O2} Determine how driven turbulence differes from that in undisturbed SW.
    \end{itemize}   
\end{itemize}

The HS goals and objectives in turn define the observatory and instrument requirements, detailed in Secs. \ref{sec:requirements} and \ref{sec:instruments}.

\subsection{G1: Reveal the 3D spatial structure and dynamics of turbulence in a weakly collisional plasma}
\label{ssec:G1}

Most of our limited present understanding of turbulence is based on single point observations.
Clusters of four spacecraft provide improvements by exploring processes occurring at a single size scale at a single time. 
As any three points define a plane, extraction of non-coplanar 3D information (such as curls or gradients) requires four points and appropriate analysis methods \cite{Paschmann:1998,Paschmann:2008}. 
However, turbulence is fundamentally multiscale; HS for the first time simultaneously explores the dynamics of processes at multiple size scales.

\subsubsection{G1O1: Reveal how turbulent energy transfers in the typical SW plasma as a function of scale and time}
\label{sssec:G1O1}

Using the undisturbed SW as a natural laboratory, with typical plasma parameters, HS measures fluctuations in the plasma velocity and density ($\delta \V{v}$ and $\delta n$) and magnetic field ($\delta \V{B}$) at MHD to sub-ion scales simultaneously using the instrument suite described in Sec.~\ref{sec:instruments}.
These data reveal how turbulent energy is distributed and transferred as a function of space and time. 
Turbulent fluctuations are affected by local magnetic fields \cite{Iroshnikov:1963,Kraichnan:1965,Schekochihin:2009,Matthaeus:1990}, so we must characterize SW turbulence relative to the local IMF direction. 
Such studies have been performed with data from single spacecraft; c.f. the review in Chen (2016) \cite{Chen:2016}, and necessarily rely upon the assumption of essentially frozen turbulence structures, an approximation known as the Taylor hypothesis \cite{Taylor:1938,Fredricks:1976,Osman:2007,Klein:2014b} that neglects temporal variations and can infer only 1D variation along the SW flow direction. 
These studies also frequently assume that the turbulence is insensitive to the angle between the SW velocity and the magnetic field, using variations in $\theta_{vB}$ to study the functional dependence of the turbulence on the angle between the wavevector and magnetic field $\theta_{kB}$.
Recent work \cite{Woodham:2021b} suggests that this assumption may not be valid; verifying this claim will require sampling the turbulent structures both along and transverse to the magnetic field direction simultaneously, a measurement that HS is designed to produce.

With HS, the Taylor hypothesis can be directly evaluated. 
Spectral information is also available from proven analysis techniques (Sec.~\ref{sec:analysis.methods}) such as 2-point correlations, structure functions, space-time correlations, and cascade rate analysis \cite{Matthaeus:1982c,Horbury:2012,Matthaeus:1990,Hamilton:2008,Chen:2011a,Horbury:2008,Mallet:2016}, from which it is possible to extract information about 3D spectral structure \cite{Osman:2011b,Hamilton:2008,Bieber:1996} and its intrinsic, scale-dependent decorrelation times.
These techniques frequently use measurements of the velocity and magnetic fields directly, or the Elsasser variables $(\V{z}^\pm = \delta \V{v} \pm \delta \V{b})$ \cite{Elsasser:1950} in which the magnetic field is expressed in \Alfven (velocity) units
$(\delta \V{b} = \delta \V{B}/\sqrt{\mu_0 n_p m_p})$ and $\delta$ indicates the use of a fluctuating quantity.

A prominent example of the use of Elsasser variables is the MHD $3^{\textrm{rd}}$-order law 
\begin{equation}
\nabla \cdot \left<\Delta \V{z}^{\mp}\vert\Delta \V{z}^{\pm}\vert ^2\right> 
= -4 \epsilon^{\pm},
\end{equation}
an analytic result involving spatial increments $\Delta \V{x}$ of the Elsasser fields $\Delta \V{z}^{\pm} = \V{z}^{\pm}(\V{x}+\Delta \V{x})-\V{z}^\pm(\V{x})$, and $\left<...\right>$ denotes ensemble average.  
This relation can be used to determine the energy cascade rate associated with the forward and backward Elsasser fields $\epsilon^{\pm}$ \cite{Osman:2011b}. 
Formally, this requires knowledge of 3D anisotropies. 
Previous studies have usually made assumptions about isotropy \cite{MacBride:2005,MacBride:2008,Stawarz:2009,Coburn:2012,Hadid:2017} or only measured $\epsilon$ over limited range of scales \cite{Bandyopadhyay:2018}. 
HS can implement the isotropic form at all nine spacecraft, but also can integrate the 3D form of the $3^{\textrm{rd}}$-order law at several scales simultaneously, making use of all 36 spacecraft pairs to compute the 2-point spatial increments. 
HS provides simultaneous 3D multipoint knowledge needed to infer spatial gradients contained in the $3^{\textrm{rd}}$-order equation, quantifying directly those key terms for the first time, bypassing simplifying assumptions about isotropy, to measure cross-scale energy transfer rates definitively.

No comprehensive observational evidence exists to distinguish between proposed theories of turbulent energy transfer. 
A review of such theories can be found in NAS 2020 Plasma Decadal Panel white papers \cite{Klein:2019:WP,Matthaeus:2019:WP,TenBarge:2019:WP} and other reviews \cite{Schekochihin:2009,Oughton:2017}. 
Candidate energy transfer processes are related to relevant dynamical timescales that include wave propagation, random and coherent sweeping of small structures by larger structures, and nonlinear wave distortion \cite{Orszag:1972,Tennekes:1975,Nelkin:1990,Sanda:1992,Servidio:2011}.
Numerical simulations provide insights regarding which of these are important but results remain inconclusive due to fundamental limitations associated with the necessary trade offs between the volume of space simulated and the physical processes included in the equations evolved.
HS provides observations to distinguish and refine our understanding of the relevance of these processes. 

\subsubsection{G1O2: Reveal how turbulent cascade of energy varies with background parameters in different SW environments}
\label{sssec:G1O2}

Turbulence and plasma conditions in fast and slow SW differ systematically in terms of density, temperature anisotropy, and collisional age \cite{Belcher:1971,Dasso:2005,MacBride:2005,MacBride:2008,Borovsky:2019,Kasper:2008}. 
Slow SW turbulence is more highly variable in nature than the fast SW \cite{Dasso:2005} and due to its longer transit from the Sun, has more time to evolve toward a fully developed state. 
These differences have been assessed in limited fashion with single-point (\cite{Breech:2008,Vech:2017}, e.g., Wind, Voyager) and single-scale \cite{Bandyopadhyay:2018} (e.g., MMS) measurements.
The varying SW speed is also associated with variations in proton number density, temperature, alpha particle density, and IMF strength. 
Plasma $\beta=8 \pi n k_b T/B^2$, the ratio of thermal to magnetic pressure, a particularly important regulator of plasma processes \cite{Chen:2014}, and power imbalances (such as cross helicity $\sigma_C$ and residual energy $\sigma_R$, \cite{Wicks:2013}), are also highly variable in the SW. 
These parameters influence the underlying energy cascade from MHD to sub-ion scales. 
HS targets to study the impact of this variability on the dynamics of the turbulence.

\subsubsection{G1O3: Quantify transfer of turbulent energy between fields, flows, and proton heat}
\label{sssec:G1O3}

Dissipation of turbulence is one of the most important factors influencing heating and particle energization in the universe. 
Consequently, our goal of investigating energy transfer must include how the cascade heats protons. 
Protons are of primary importance as they are the dominant species in terms of both mass and momentum. 
How and how much energy is delivered to protons via dissipation processes determines the overall partitioning of energy across all species. 
Primary candidate mechanisms include: ICWs and cyclotron resonances \cite{Hollweg:2002}; Landau damping \cite{TenBarge:2013a}; stochastic heating by large amplitude turbulent fluctuations \cite{Chandran:2010a}; and energization through intermittent structures, including magnetic reconnection \cite{Dmitruk:2004} and trapping in secondary magnetic islands \cite{Ambrosiano:1988}.

Current observations do not provide clarity. 
For example, intense ICWs are commonly observed during times when plasma instabilities are present \cite{Gary:2016} in extended “storms” during quiet SW and radial IMF \cite{Jian:2014}. 
Because ICWs are capable of substantial heating of SW ions, it is important to understand exactly how often they occur. 
ICWs may be omnipresent but can only be detected by a single spacecraft when the SW flow is aligned with the local IMF (i.e., radial field configurations). 
Applying methods such as the wave telescope technique to HS observations, Sec.~\ref{sec:analysis.methods}, will identify ICWs when the IMF is not radial, thus establishing definitively whether ICWs are always present or not.

All aforementioned mechanisms occur at ion time and length scales and create characteristic signatures in underlying proton velocity distribution functions (VDFs); each mechanism deposits differing fractions of energy to the protons \cite{Chandran:2010a,He:2015,Matthaeus:2016a}. 
The absence or presence of these signatures reveals which dissipation pathways operate; their relative strengths quantify their relative importance.
Measurements of proton temperature at ion heating time scales allows HS to quantify proton heating directly. 
One analysis method, colloquially referred to as 'PiD', makes use of the measured pressure tensor $\Pi_{ij}$ and flow gradients $S_{ij}=\nabla_i \V{u}_j$ to compute the full pressure-strain interactions $\Pi:S$ which is the rate of production of proton internal energy \cite{Yang:2019}. 
These methods are enabled in HS by simultaneous measurement of proton distribution functions and 3D multiscale turbulence, a \change{combined} capability lacking in all previous missions. 
HS will allow us to directly quantify relationships between the distribution of turbulence fluctuations and transformation into proton heat.

\subsubsection{G1O4: Identify thermodynamic impacts of intermittent structures on protons}
\label{sssec:G1O4}

Intermittency is a universal property of turbulence in which dissipation processes concentrate into small fractions of available volumes \cite{Horbury:1997,Matthaeus:2015}, giving rise to current sheet structures. 
Such structures have been studied in numerical simulations \cite{Karimabadi:2013} and \textit{in situ} observations \cite{Osman:2011}.
Intermittency dramatically impacts how turbulence heats plasma \cite{Mallet:2019}. 
Cluster and MMS pioneered the ability to resolve thin structures with 4-point curlometer and gradient techniques \cite{Dunlop:2002a,Dunlop:2002b}. 
While revolutionary, such techniques probe only a single scale at a time. 
HS provides combinatorically more spacecraft groupings and simultaneous access to multiple scales, tremendously expanding 3D anisotropic measures of intermittency with well-developed analysis tools, as described in Sec.~\ref{sec:analysis.methods}. 

By measuring intermittency of turbulent fluctuations at inertial and ion scales simultaneously, HS differentiates between models of nonlinear coupling, that predict enhanced amplitudes of Elsasser fluctuations $\delta z$ at small scales compared to a scale-independent normal distribution of amplitudes.
HS also resolves the intermediate scales to provide further differentiation.

\subsection{G2: Ascertain the mutual impacts of turbulence, variability, and boundaries near large scale structures}
\label{ssec:G2}

While undisturbed SW is a pristine environment, disturbed SW occurs from impacts of either large scale structures of solar or heliospheric origin or Earth’s magnetosphere and provides different environments to explore. 
Impacts are mutual: turbulence can impact large-scale structures and boundaries and those same structures can in turn change the nature of turbulence. 
Goal Two focuses on these mutual interactions.

\subsubsection{G2O1: Determine how SW turbulence affects and is affected by large-scale structures such as CMEs and CIRs}
\label{sssec:G2O1}

Passage of interplanetary coronal mass ejections (CMEs)\cite{Jian:2018} or corotating interaction regions (CIRs)\cite{Goldstein:1984,Jian:2019} disturb the SW from its pristine state.
These large-scale features re-inject energy and thus modify SW turbulence. 
While turbulence levels are reduced within CME structures, HS enables 3D characterization of this (possibly weak) low-plasma $\beta$ turbulence contained within a large scale force-free structure.
Near CMEs, driven turbulence departs significantly from that of the pristine SW; HS will diagnose 3D turbulence modifications associated with diffusive shock acceleration \cite{LeRoux:2015} near fast CMEs and waves driven by the CME's propagation\cite{Zhao:2021a}.
Passage of both CMEs and CIRs through pristine SW turbulence allows us to explore differences in these environments, enabling us to determine when and how specific energy transfer and heating processes become important.

\subsubsection{G2O2: Determine how driven turbulence differs from that in undisturbed SW}
\label{sssec:G2O2}

The terrestrial bow shock, foreshock, and magnetosheath are permeated with magnetic and plasma fluctuations, strongly driving and modifying the turbulent spectrum across inertial and dissipation scales both in amplitude and shape\cite{Chen:2019:WP}.
These regions represent parameter regimes not accessible in pristine SW.
The dynamics in these locations are significantly different; for example, ions reflected off the bow shock can lead to the self-generation of turbulence, which takes the form of non-linear wave penetrating into the inner magnetosphere \cite{Takahashi:2016}, while at the shock, turbulence generates high-speed jets that regularly impact the magnetopause, resulting in dayside reconnection \cite{Hietala:2018}.
Turbulence is also seen to drive magnetic reconnection in these regions \cite{Retino:2007}.
Finally, magnetospheric regions can be turbulent \cite{Chasapis:2017,Chasapis:2020,Bandyopadhyay:2020a,Bandyopadhyay:2020b}, but of a different nature (e.g. magnetically dominated) \cite{Maruca:2018}. 
Objective Two explores this variety of accessible systems to compare how driven environments differ from pristine SW.

\section{HelioSwarm Observatory Design}
\label{sec:requirements}

The specific design of the HS Observatory is driven by decades of measurements from near-Earth plasmas of characteristic length and time scales as well as derived dimensionless parameters that are predicted to govern the behavior of magnetized turbulence.

\subsection{Quantities to be Measured}
\label{ssec:requirements.mr}

As discussed in Sec.~\ref{sssec:G1O1}, the primitive variables that describe magnetized turbulence at MHD scales are the Elsasser variables \cite{Elsasser:1950} composed of magnetic fields and particle densities and velocities.
G1O1 requires measurements of the IMF, SW proton density, and SW velocity.
It must do so in undisturbed, most-probable SW for which the range of proton densities is $1.6$
to $20$ cm$^{-3}$ and magnetic field can be as large as 25 nT , but typically
larger than 2.6 nT (at the 90\% occurrence rate) \cite{Wilson:2018,Klein:2019}.
To resolve at the lowest typical field strength, we require 10\% resolution (0.26 nT), corresponding to 0.15 nT per axis. 
Such measurements allow construction of Elsasser variables, needed for magnetized turbulence analysis at each measurement location.

Measurements of the SW proton density, velocity and IMF must be made at multiple points in 3D encompassing the turbulent cascade during average SW conditions, within large scale structure analysis intervals---equivalent to approximately one hour long continuous observations--- at cadences, time knowledge, and sensitivities required to resolve and align SW and IMF variations down to sub-ion scales.

\subsection{Spatial Resolution}
\label{ssec:space.data}

\begin{figure}
    \centering
                \subfloat{\includegraphics[height=28ex]{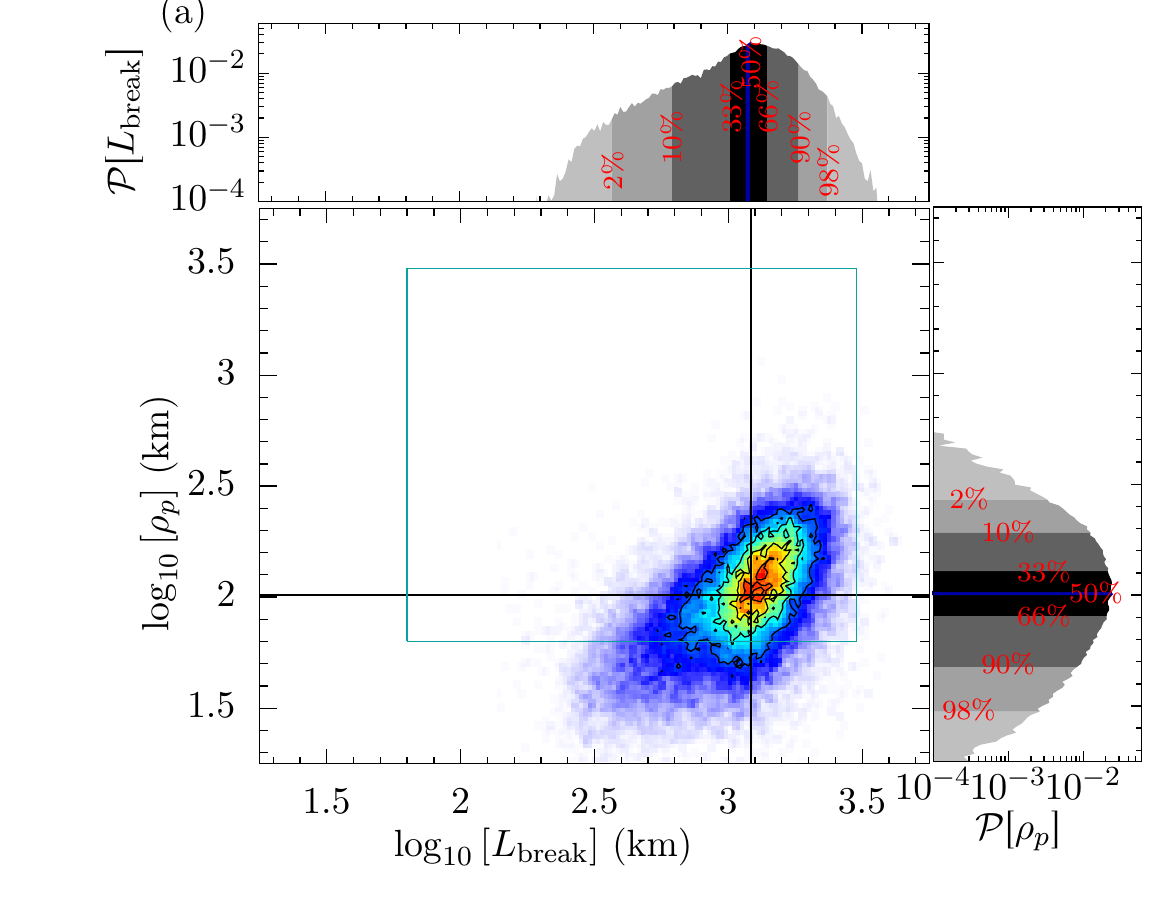}}
    \subfloat{\includegraphics[height=28ex]{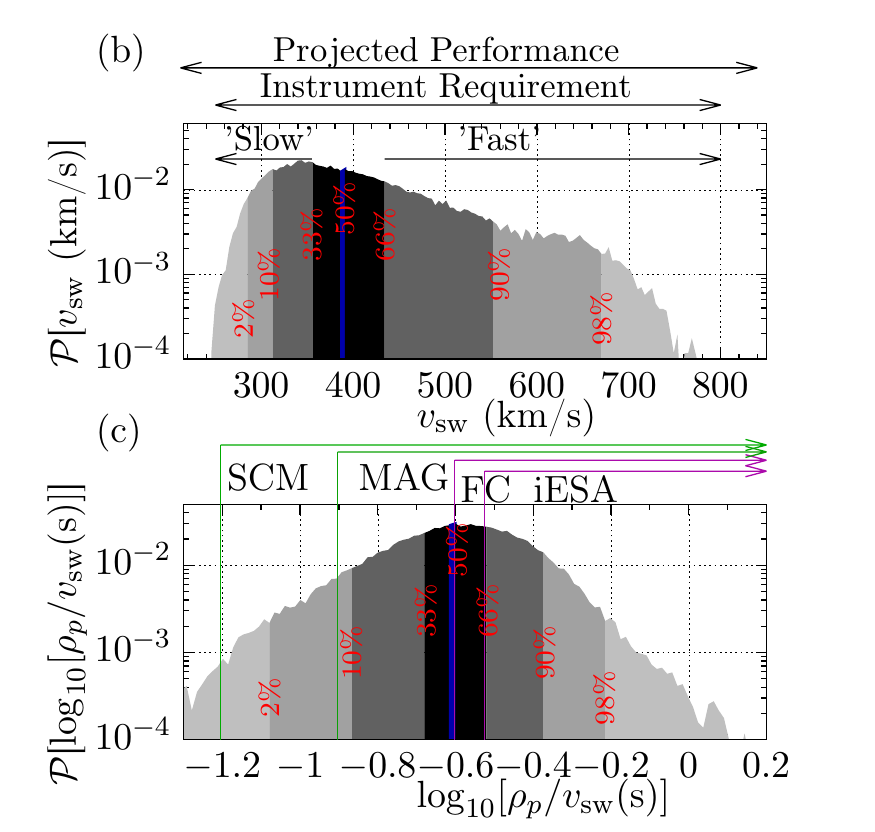}}

    \caption{(a) Joint PDF of proton gyroradius $\rho_p$ and spectral break scale $L_{\textrm{break}}$ as measured by the Wind spacecraft at Earth's L1 point \cite{Wilson:2018,Klein:2019}.
    HS's baseline separations between spacecraft will cover from 3000 km to 50 km (blue box), allowing the observatory to simultaneously measure MHD, transition, and sub-ion physical processes in 85\% of the pristine SW.
    (b) PDF of SW velocity drawn from the same database, compared to FC instrumental requirements and project performance, illustrating that HS will capture both typical and extreme proton velocities.
    (c) PDF of the advected SW ion timescale $\rho_p/v_{\textrm{sw}}$, compared to HS instrument cadences, demonstrating that HS will resolve the IMF past ion-scales in nearly all the SW, and resolve both the ion-scale plasma processes in typical SW conditions.
    In all panels, the red numbers indicate the percentile of the cumulative distribution below the given value.
    }
    \label{fig:joint_PDF}
\end{figure}

To measure the multiscale nature of turbulence, HS's baseline separations between the nine spacecraft are designed to simultaneously span MHD scales and ion kinetic scales, enabling the simultaneous resolution of MHD and sub-ion processes and the transition between these scales, exemplified by the observed spectral break \cite{Goldstein:1994,Leamon:1999,Hamilton:2008,Chen:2014,Vech:2017,Vech:2018,Woodham:2018} (see also Figure~\ref{fig:spectra}).
Values for these physical dimensions are empirically known from decades of SW observations\cite{Borovsky:2019,Wilson:2018,Woodham:2018,Klein:2019}. 
Figure~\ref{fig:joint_PDF} shows the joint probability distribution function (PDF) of the proton gyroscale $\rho_p$ and spectral break scale $L_{\textrm{break}} = v_{\textrm{sw}} /f_{\textrm{break}}$.
These observations define three ranges: MHD scales at $>1200$ km; transition scales between 100 and 1200 km; and sub-ion structures $<100$ km. 
HS's baseline requirements are established to resolve these characteristic scales simultaneously in 85\% of the pristine SW, enabling the Observatory to “encompass the turbulent cascade.”

\subsection{Temporal Resolution}
\label{ssec:time.data}

The Observatory measurement cadence and timing knowledge provide the temporal resolution necessary to resolve advected SW structures.
This analysis requires measuring at time cadences from MHD scales down to the sub-ion scales.

Given the observed distribution of SW velocities, see Figure~\ref{fig:joint_PDF}b,
we can calculate the ratio of the proton gyroscale $\rho_p$ to $v_{\textrm{SW}}$ to construct an advected proton timescale, Figure~\ref{fig:joint_PDF}c, which plots the observed distribution against instrumental measurement rates.
The fluxgate magnetometer (MAG) measures at 16 samples per second (Sps) overlapping with the searchcoil magnetometer (SCM, at 32 Sps) providing continuous coverage of larger and/or more slowly advecting structures, while also  resolving ion-scale structures traveling at the fastest $ v_{\textrm{SW}} (\sim 800$ km/s); 
The proton density ($n$) and velocity ($v$) are measured by Faraday cups (FCs) at a rate of 8 Sps, resolving ion scale structures ($\sim 100$ km) traveling at typical speeds (400 km/s);
Measurements of the proton temperature by the ion electrostatic analyzer (iESA) provide the necessary context for the kind of turbulence HS is embedded in, with sufficient temporal resolution to resolve changes in proton velocity distributions to help determine the energy transfer processes associated with ion scale structure.

In order to resolve characteristic SW wave propagation directions across multiple points, HS requires post-facto, relative pairwise separation knowledge of 10\% the separation distances. 
\change{Timing requirements are driven by applying analysis methods described in Section~\ref{sec:analysis.methods} to synthetic data combined with models for temporal uncertainty.}

\subsection{Observatory Stability}
\label{ssec:stability.data}

\begin{figure}
    \centering
         \includegraphics[width=\columnwidth]{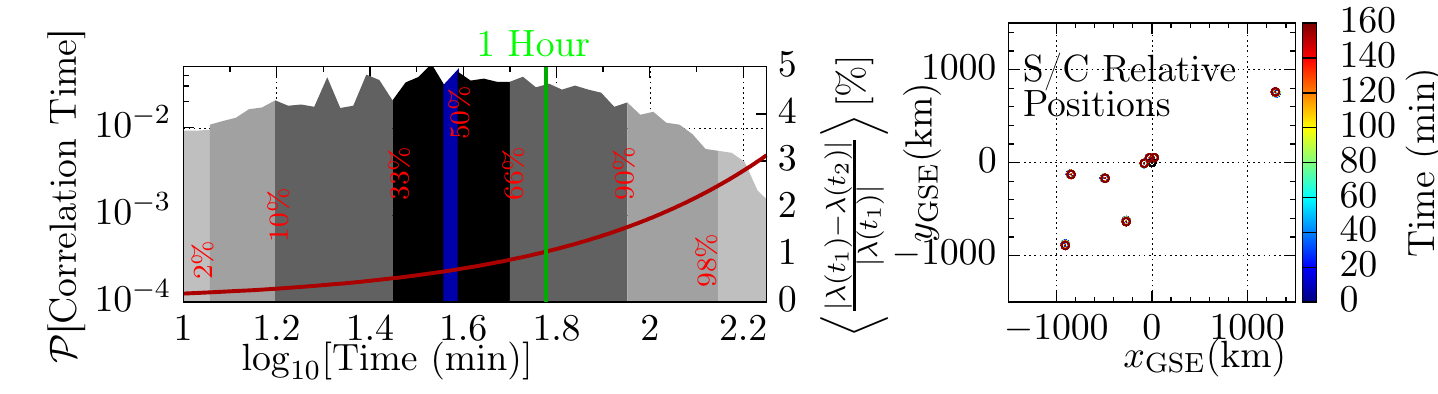}
    \caption{PDF of correlation time measured by ACE (left panel) \cite{Isaacs:2015} compared to the average change in HS DRM baseline separation magnitudes as a function of time (red line). 
    At right, the $x_{\textrm{GSE}}-y_{\textrm{GSE}}$ projection of the evolving Node positions relative to the Hub, with color indicating time since apogee; the significant overlap in positions illustrate the relative stability of the observatory configuration.} 
    \label{fig:correlation}
\end{figure}

Simultaneous statistical analysis of turbulence (e.g. Sec.~\ref{ssec:analysis.msa}) requires not only separations spanning the previous described spatial scales but also samples taken over long enough periods of time to capture the nonlinear reshaping of the underlying structures.
One can calculate the correlation time scale $\tau$ by determining the time lag necessary to reduce an autocorrelation of some measured quantity $F$ by $1/e$ from it's zero-lag value
\begin{equation}
    A\left[F(t),F(t+\tau)\right]=\left<F(t)F(t+\tau) \right>=\left<F(t)F(t) \right>/e,
\end{equation}
where $<...>$ denotes an appropriate ensemble average.
Analysis performed on intervals measured within a correlation time are effectively sampling the same population of turbulent fluctuations, and thus can be combined to study the statistical properties of that plasma. 
Observations of the correlation time scale in the SW\cite{Isaacs:2015,Smith:2018}, illustrated in Figure~\ref{fig:correlation}, typically find it ranges from tens of minutes to approximately an hour.
This duration of SW data provides robust turbulence analysis yet is short enough to effectively sample the same parcel of SW.
These observations drive the timescales over which the observatory spacecraft separations need to be constant, a requirement the HS Design Reference Mission (DRM) satisfies,
enabling the accrual of usable intervals for the application of analysis approaches outlined in Sec.~\ref{sec:analysis.methods}; the average relative change in the vector baselines increases slowly in time (red line), reaching 0.7\% at 60 minutes and 1.5\% at 120 minutes.

\subsection{Spatial Configurations}
\label{ssec:configurations.data}

\begin{table}[]
    \centering
    \begin{tabular}{|l|ccccc|c|cc|}
    \hline
    \# Spacecraft & 4 &	5&	6&	7&	8&	9&	10&	11 \\
    \hline
Baselines & 6&	10&	15&	21&	28&	36&	45&	55\\
Tetrahedra & 1&	5&	15&	35&	70&	126&	210&	330 \\
Polyhedra & 1&	6&	22&	64&	163&	382&	848&	1816 \\
    \hline
    \end{tabular}
    \caption{Number of baselines, tetrahedral, and polyhedral (with at least four vertices) configurations that can be constructed from $N$ Spacecraft.}
    \label{tab:baselines}
\end{table}

\begin{figure}[H]
{\includegraphics[width=\textwidth]{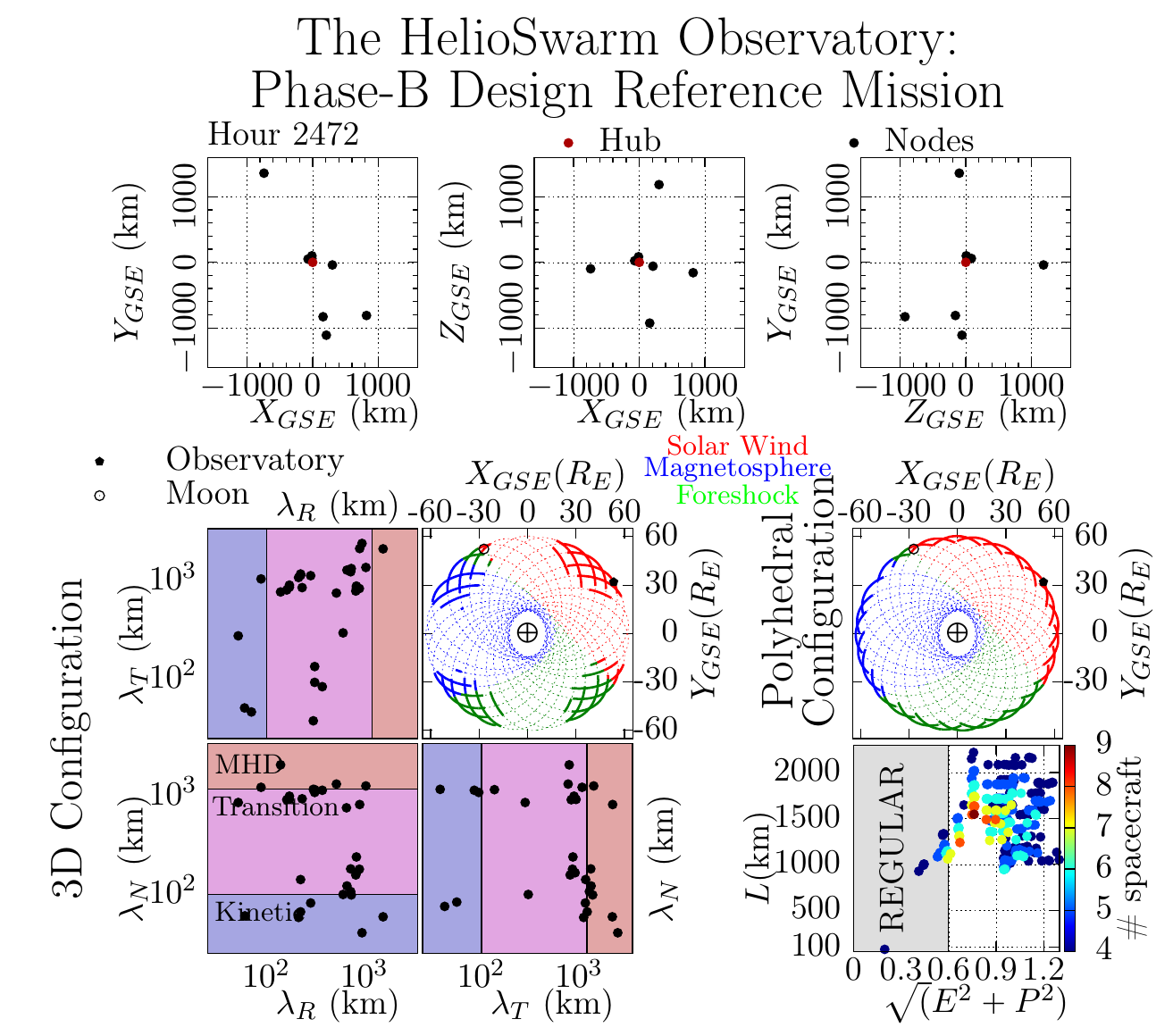}}
    \caption{Summary plot of HelioSwarm Observatory Phase-B DRM positions and separations. \textit{Top Row} 
    Relative positions between the Hub (red) and eight Nodes (black) projected into the GSE co\"ordinate system at hour 2472 from the DRM.
    \textit{Bottom Left} 
    Projected vector components of the 36 inter-spacecraft baseline separations (black dots) demonstrate coverage of MHD and ion-kinetic scales, as well as the transition region in-between. 
    The lunar resonant orbit of the observatory (black dot) in the GSE co\"ordinate system is shown as colored lines in the upper-right inset, with the moon's location (open circle) included to illustrate scale. 
    Times with orthogonal coverage over all three scales, highlighted in color, arise in the pristine SW (red lines), the magnetically connected SW (green) and the magnetosphere/magnetosheath (blue). 
    \textit{Bottom Right}
    The size and geometric configurations of the polyhedra constructed by spacecraft subsets of the HelioSwarm observatory.
    The number of spacecraft is indicated by color, while the size of the polyhedra $L$ and its regularity (the RMS of the elongation $E$ and planarity $P$) are indicated on the ordinate and abscissa respectively.
    The times when there are at least two regular polyhedra with characteristic sizes more than a factor of three different are indicted in the upper inset, using the same color scheme as the 3D Configuration inset.
    As quantified in Table~\ref{tab:residence_time}, due to the high eccentricity of the orbit, the Observatory samples these regions near apogee for a substantial fraction of the orbit period.
    A video of the HS DRM Geometries throughout the Science Phase is available in Online Resource 2.
    }
    \label{fig:example}
\end{figure}

In conjunction with spatial separation requirements, the application of the analysis approaches in Sec.~\ref{sec:analysis.methods} require specific spatial configurations. 
Given $N$ spacecraft, there are $N(N-1)/2$ distinct pair-wise baseline separations.
Similarly, for $N$ spacecraft, one can construct ${N \choose 4} = \frac{N!}{4!(N-4)!}$ unique tetrahedral configurations, or $\sum_{i=4}^{N} {N \choose i}$ polyhedral configurations with at least four vertices. The number of baselines, tetrahedra, and polyhedra are tabulated as a function of the number of spacecraft in Table~\ref{tab:baselines}. 
The orientation and geometries of these configurations have been carefully tailored so that they span the appropriate size-scales and directions to address the mission objectives, as discussed in the following subsections and illustrated in Fig.~\ref{fig:example}.
Determining when the HS Observatory satisfies these configurational requirements is characterized in 1-hour units, during which baseline separations are effectively constant, see Fig.~\ref{fig:correlation}.
The number of hours satisfying these requirements are laid out in Table~\ref{tab:residence_time}.

\subsubsection{3D Configurations}
\label{sssec:3D}


To calculate cascade rates, correlation scales, and structure functions to characterize the multiscale and 3D nature of turbulence, 
the 36 unique baselines between HS’s nine spacecraft have vector components spanning three orthogonal directions along, transverse, and normal to the Earth-Sun line (Radial, Tangential, Normal (RTN) co\"ordinates) with amplitudes covering MHD, transition, and sub-ion scales, while simultaneously the magnitudes of the baseline vectors also span these three ranges of scales.
These \textit{3D configurations}, illustrated in Figure~\ref{fig:example}, resolve variations along and across the local magnetic field and flow directions, necessary for verifying theories of anisotropic turbulent transfer and distribution of energy.

\subsubsection{Polyhedral Configurations}
\label{sssec:poly}

The polyhedral configuration is satisfied when at least two of HS's constituent polyhedra, each satisfying $\sqrt{E^2 +P^2} \leq 0.6$, have at least a factor of three difference in L, simultaneously measuring the spatial structure of turbulence at multiple scales.

\begin{figure}
    \centering
    \includegraphics[width=0.5\columnwidth]{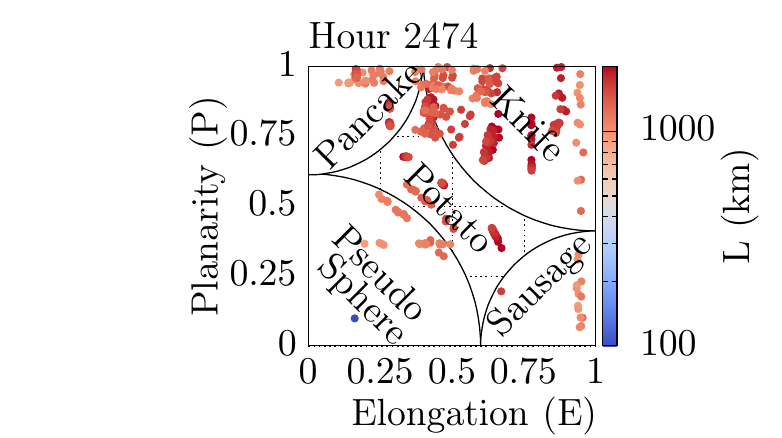}
    \caption{The distribution of planarity (P), elongation (E), and characteristic size (L, \change{with blue and red representing the smallest and largest scales respectively}) of all 382 polyhedra with at least 4 vertices for the HS DRM at an arbitrarily selected hour 2474. 
    The different regions in E-P space are labeled to characterize the geometries of these polyhedra.
    The Observatory trajectories are designed to have multiple pseudo-spherical polyhedra with significantly different sizes to enable measurements of spatial structures at MHD- and ion-scales simultaneously.}
    \label{fig:poly_config}
\end{figure}

HS configurations are also designed for multi-point analysis techniques which determine spatial gradients and distributions of power (e.g., wave-telescope, curlometer, and related gradient analysis techniques \cite{Paschmann:1998,Paschmann:2008}. 
Spatial gradient methods require the SC be arranged in a quasi-regular fashion, occupying vertices of pseudo-spherical polyhedra. 
One can characterize the geometry of these polyhedra by calculating the eigenvectors of the volumetric tensor
\begin{equation}
    \underline{\underline{R}} = \frac{1}{N} \sum_{\alpha=1}^N\left(\V{r}_\alpha-\V{r}_b\right)\left(\V{r}_\alpha-\V{r}_b\right)^T
\end{equation}
where $\V{r}_b=\frac{1}{N}\sum_{\alpha=1}^N\V{r}_\alpha$ is the mesocenter of the configuration, and $\V{r}_\alpha$ represents the positions of the individual SC.
The square roots of the three eigenvalues of $\underline{\underline{R}}$ represent the major, middle, and minor semiaxes of the configuration, $a$, $b$, and $c$.
These values can be interpreted directly by defining a characteristic size $L=2 a$, as well as the elongation $E=1-\frac{b}{a}$ and planarity $P=1-\frac{c}{b}$.
Figure~\ref{fig:poly_config} illustrates the distribution of polyhedra from a single hour of the HS Observatory configuration.
Polyhedra with small elongation E and planarity P, $\sqrt{E^2 +P^2} \leq 0.6$, can be used to accurately measure structure of sizes on the order of the characteristic size $L$ \cite{Sahraoui:2010b,Roberts:2015b}. 
HS's 9 spacecraft produce 382 polyhedra with at least 4 vertices, the minimum needed for 3D analysis techniques, and at many different scales. 
Additionally, HS has configurations where at least two pseudo-spherical polyhedra exist with at least a 3:1 ratio in L. 
These formations, referred to as polyhedral configurations, simultaneously measure spatial structure of turbulence at multiple scales.

\subsection{Observatory Orbits}
\label{ssec:orbits}

The HS Observatory accesses the near-Earth regions of interest with a 2-week, lunar-resonant, high Earth orbit (HEO) \cite{Plice:2019,Levinson-Muth:2021a,Levinson-Muth:2021b,Levinson-Muth:2022}.
The HS Observatory design and onboard propulsion produce inter-spacecraft separations both along and across the Sun-Earth line. 
The Nodes perform routine trim maneuvers to maintain customized configurations that satisfy the 3D and Polyhedral requirements over the mission lifetime. 
Since the science orbit is nearly inertially fixed (with a low rate of apsidal precession), the apogee rotates through the SW, foreshock, and magnetosphere-dominated regions as the Earth completes a single orbit of the Sun.
This progression allows the Observatory to sample the pristine SW and regions of strongly driven turbulence during the 12-month Science Phase, addressing both G1 and G2. 




Given an empirical model for the extent of the magnetosphere \cite{Formisano:1979} and the average orientation of the IMF combined with the phase B DRM trajectories, the HS Observatory spends thousands of hours in the required near-Earth regions of interest, with hundreds of hours in both of the required spatial configurations in each region; see Table~\ref{tab:residence_time} for summary of hours and Figure~\ref{fig:example} for an illustration of the residence time in the regions.
Measurements from all instruments are recorded throughout the orbits outside of thruster operations, eclipses, and calibration activities and transmitted regardless of the Observatory configuration.

\begin{table}[]
    \centering
    \begin{tabular}{|c|c|c|c|c|}
    \hline
    Phase B DRM; LRD 2028 & Fig.~\ref{fig:example} & Total & 3D & Polyhedral \\
    \hline
    Solar Wind & Red & 2881 & 777 & 1068 \\
    Foreshock & Green & 2470 & 977 & 852 \\
    Magnetosphere/Magnetosheath & Blue & 3149 & 650 & 639 \\
    \hline
    Science Phase & & 8850 & 2404 & 2559 \\
    \hline
    \hline
    & Objective & Total & 3D & Polyhedral \\
    \hline
    Pristine SW & G1O1 & 2015 & 544 & 747 \\
    Extreme SW & G1O2 & 58 & 16 & 21 \\
    SW w/ Large Scale Structure & G2O1 & 546 & 147 & 202 \\
    Strongly Driven Turbulence & G2O2 & 5619 & 1627 & 1491 \\
    \hline
    \end{tabular}
    \caption{HS measures thousands of hours in targeted
near-Earth regions of space, with hundreds of hours in optimal
polyhedral (Sec.~\ref{sssec:poly}) and 3D configurations (Sec.~\ref{sssec:3D}) for the application of analysis approaches outlined in Sec.~\ref{sec:analysis.methods}, providing measurements to advance the understanding of turbulence in typical (G1O1,G2O2) uncommon (G2O1) and extreme (G1O2) plasma conditions.}
    \label{tab:residence_time}
\end{table}

\subsection{Mission Duration}
\label{ssec:duration.data}

As discussed in Sec.~\ref{sssec:G1O2}, SW parameters drive the behavior of turbulence, and more extreme values of these parameters are useful for distinguishing competing theories. 
To establish the minimum number of hours needed for HS science data sufficiency, we note that $\sim 10$ hours in extremely high ($\geq 10$) and low ($\leq 0.1$) plasma $\beta$ enabled strong characterization of the spectral break, differentiating between predicted dissipation mechanisms \cite{Chen:2014}. 
Using this assessment, the required number of hours of observation for the Baseline Mission was developed by analyzing two decades of Wind \cite{Wilson:2018,Klein:2019} data to ensure that we would adequately sample the full range of SW variability. 
Our methodology was to generate PDFs of parameters controlling turbulent behavior, e.g. SW speed ($v_{\textrm{SW}}$), plasma beta ($\beta$), proton temperature ($T_p$), balance of power between Sunward and anti-Sunward propagating fluctuations (cross helicity, $\sigma_C$ \cite{Parashar:2018}), difference between kinetic and magnetic energy (residual energy $\sigma_R$ \cite{Wicks:2013}), SW collisionality (Coulomb number $N_C$ \cite{Neugebauer:1976,Kasper:2017}), and Alfv\'en Mach number ($v_{\textrm{SW}}/v_A$). 
Obtaining $\sim 10$ hours of measuring turbulence at relatively large and small values of these parameters determines the overall requirements for the number of hours in polyhedral and 3-D configurations; from the widths of the parameter PDFs, we determined that measuring 500 (100) hours in the 3D (polyhedral) configuration in the pristine SW, which then result in HS measuring 10 (2) hours of turbulence with extreme parameters both higher than the 98th percentile and lower than 2nd percentile of those values (corresponding to $\beta \sim 0.1$ and $\beta\sim10$ \cite{Chen:2014}), a sufficient number of intervals at the very most extreme parameters to accomplish Mission science. 
Magnetosheath plasmas that will also be measured by HS typically have even higher values of $\beta_{\parallel,p}$\cite{Maruca:2018}. 
Measurements of these extreme intervals allow for the identification of different turbulent processes that are preferred in different parameter regimes, and will also be useful for providing accessible analogies to astrophysical systems where the thermal pressure dominates, e.g., the ISM ($\beta_{\parallel,p}\gtrsim 10$) or accretion disks ($\beta_{\parallel,p}\gtrsim 1$).

Large scale structures (LSS) generated by the Sun– e.g., CMEs, or produced as the SW propagates, e.g., CIRs can drive different kinds of turbulence compared to SW w/o LSS. 
By using \textit{in situ} SW measurements of these structures over the last two solar cycles \cite{Jian:2018,Jian:2019}, we calculate the filling fraction during the 12-month Science Phase of these two kinds of structures
for a 2028 Launch Readiness Date (LRD) based upon equivalent phases from Solar Cycles 23 and 24; the total anticipated LSS hours for this LRD are tabulated in Table~\ref{tab:residence_time}.

The average CME filling fraction is 2.15\% (1.9\%/2.4\% in Solar Cycle 23/24) while the CIR filling fraction is 16.8\% (19.8\%/13.8\% in Solar Cycle 23/24). 
These rates correspond to 62 hours of CME observations, with 16/23 hours in 3D/polyhedral configurations and 484 hours of CIR measurements, with 131/179 hours in 3D/polyhedral configurations. 
We have repeated this exercise for other LRDs, and found that regardless of launch date, there will be a sufficient number of hours of observed CMEs and CIRs to provide data to bring closure to G2O1.


\subsection{Resilience, Redundancy, and Robustness of Multi-satellite Observatory}
Multi-SC swarm design offers innovations in flexibility and reconfiguration of the observatory.
Orbital mechanics forces create continuously evolving relative positions among the 9 SC in HS.
With known exceptions, the nominal swarm configuration has redundancy in most of the 3D baselines and tetrahedral vertices and accrues successful hours of science data collection well above the requirements.

Robustness above required performance and redundancy in spatial configurations create resilience in the event of contingencies.
For the case of the loss of any one (or two) Nodes, \change{the required number of hours in both configurations} can be achieved within the duration of the 12-month Science Phase through a repositioning of the remaining Nodes to construct the configurations for which sufficient hours have not been achieved.


\subsection{Place within the Heliospheric System Observatory}
\label{ssec:HSO}

HS stands alone, but would also be part of the Heliospheric System Observatory (HSO) which provides additional opportunities for joint mission studies. 
Parker Solar Probe (PSP) \cite{Fox:2015} and Solar Orbiter (SolO) \cite{Muller:2013} are making high-cadence plasma and IMF measurements of the innermost heliosphere.
These inner-heliospheric missions provide only single point measurements, but these inform limits on injection scale structures that cascade into smaller structures as they propagate to 1 AU. 
Together with HS, and supplemented by Polarimeter to UNify the Corona and Heliosphere (PUNCH) imaging\cite{DeForest:2018}, these observations allow for estimates of 1-AU-scale evolution and radial and longitudinal gradients.
At intermediate scales ($10^6$ km), HS observations in combination with other missions in the HSO positions near the Sun-Earth L1 point (e.g., Advanced Composition Explorer (ACE)\cite{Stone:1998}, Wind\cite{Wilson:2021}, Interstellar Mapping and Acceleration Probe (IMAP)\cite{McComas:2018}, Deep Space Climate Observatory (DSCOVR)\cite{Lotoaniu:2022}) provide opportunity for long-baseline correlations and to address the long-open question of local geometries of interplanetary shocks and flux ropes. 
These same HSO missions provide additional SW composition information to augment HS alpha particle measurements.
Conjunctions with MMS\cite{Burch:2016} may also prove useful in extending the range of scales over which energy transfer and dissipation can be studied.
Finally, given that energetic particle propagation is impacted by SW turbulence, ACE, Wind, Solar Terrestrial Relations Observatory (STEREO), and IMAP energetic particle measurements can test the effect of turbulence models and mechanisms HS quantifies.
Joint study opportunities will depend on what HSO assets are operating when HS launches, but the breadth of the positions and instrumentation of missions within the HSO will enable a variety of examinations of fundamental processes at play in our Heliosphere.

\section{HelioSwarm Mission Implementation}
\label{sec:instruments}


HS was selected as a Heliophysics Division Medium Explorer (MIDEX) mission by NASA Science Mission Directorate in 2022, and is currently in the formulation phase.
MIDEX missions are affordable testbeds for flagship science, from a cost and risk implementation perspective. 
The HS hardware and operations approach are all extremely high heritage to minimize overall project risk. 

The HS architecture consists of one central Hub, an ESPA-class (EELV Secondary Payload Adapter) spacecraft provided by Northrop Grumman, and eight co-orbiting Nodes, SmallSats provided by Blue Canyon Technologies, both high heritage, 3-axis stabilized spacecraft. 
The Hub, Sec.~\ref{ssec:Hub}, carries the eight Nodes to the science orbit.
Pairs of Nodes will then separate from the Hub over four consecutive 14-day orbits. 
Each Node, Sec.~\ref{ssec:Nodes}, possesses identical instrument suites (IS) consisting of three high-heritage, high-TRL sensors optimized for HS: the Faraday Cup (FC, Sec.~\ref{ssec:FC}), provides high cadence measurements of the SW density and flow, 
the Fluxgate Magnetometer (MAG, Sec.~\ref{ssec:MAG}), and Search Coil Magnetometer (SCM, Sec.~\ref{sssec:SCM}), provide measurements of the IMF at cadences sufficient to probe fluctuations from MHD to sub-ion scales.
The Hub has the same IS as the Nodes, plus an ion Electrostatic Analyzer (iESA, Sec.~\ref{ssec:iESA}), another high-heritage, high-TRL instrument that will provide high cadence measurements of the proton and alpha particles \change{in} order to characterize the local turbulence and to quantify ion heating.
An electron Electrostatic Analyzer, Sec.~\ref{ssec:eESA}), included as a Student Collaboration Option for installation on the Hub, provides additional context for the plasma environment sampled by the HS Observatory.

\begin{table}[]
    \centering
    \begin{tabular}{|c||l|l|l|}
    \hline
    Observable & Requirement & Projected Performance & Instrument \\
    \hline
    Multi-point vector & DC to 2-Sps & DC to 16-Sps & MAG   \\
 DC IMF $\V{B}$        & $\pm 100 nT$ & $\pm 128 nT$ &  (all SC)\\
        & 0.15 nT per axis & 0.1 nT per axis & \\
    \hline
    Multi-point vector & 0.1 to 32-Sps & up to 32-Sps & SCM   \\
     AC IMF $\V{B}$        & 15/1.5 pT/$\sqrt{}$Hz at 1/10 Hz & 6/0.6 pT/$\sqrt{}$Hz at 1/10 Hz  & (all SC)\\
    \hline
    Multi-point & 0.15 s & 0.125 s & FC   \\
    proton density $n_p$    & 0.2 - 20 cm$^{-3}$ & 0.1 - 50 cm$^{-3}$ & (all SC)   \\
            & $\pm 6\%$ & $\pm 5\%$& \\
    \hline
        Multi-point & 0.15 s & 0.125 s & FC   \\
     proton velocity $\V{v}_p$& 250 - 800 km/s & 212 - 840 km/s & (all SC)\\
            & $\pm 3\%$& Accuracy $\pm 1\%$ & \\
    \hline
    Single-point  & 0.3 s & 0.15s & iESA   \\
    proton temperature $T_p$    & $10^4- 5 \times 10^5$ K & $10^4- 10^6$ K & (Hub)   \\
            & $\pm 5\%$ & $\pm 1.8\%$ & \\
    \hline
    Single-point temperature & 0.3 s & 0.15s & iESA   \\
     anisotropy $\frac{T_\perp}{T_\parallel}$    & $0.2-5$ & $0.1-10$ & (Hub) \\
            & $\pm 6\%$ & $\pm 3.4\%$ & \\
    \hline
    Single-point $\alpha$-proton  & Hourly Averages & 10 s  & iESA   \\
    density ratio $\frac{n_\alpha}{n_p}$& $0-40\%$ & $0-100\%$ & (Hub)\\
            &$\pm 10\%$ & $\pm 3.4\%$ & \\
    \hline
    \end{tabular}
    \caption{Plasma and magnetic field observables measured across the HS Observatory.
    Required cadences, ranges, and accuracy for the measurements, as well as the projected performance and the instrument that will provide the measurement are organized by column.}
    \label{tab:observables}
\end{table}

The instruments were specifically selected to be \change{both} capable of addressing the science objectives when used as an Observatory and for having high heritage to ensure the fabrication, integration, and testing approaches for the required nine copies of flight model instruments, along with their costs and schedules, would be low risk. 

\subsection{Magnetometers}
\label{ssec:mags}

HS uses a combination of flux gate (MAG) and search coil (SCM) magnetometers to measure the IMF over the required frequency range indicated by Figure~\ref{fig:spectra} (DC $\sim 3600$ s to sub-ion $< 0.15$ s).
Two different magnetometer types are required owing to sensitivities required, especially at high frequencies (15 pT/ Hz at 1 Hz and 1.5pT/ Hz at 10 Hz – see noise floors on Figure~\ref{fig:spectra}); these same sensitivities impose mission requirements for DC and AC magnetic cleanliness.
The MAG and SCM instruments overlap in frequency allowing for cross-calibration and the production of a merged data product, as has been performed for other missions \cite{Fischer:2016,Bowen:2020c}. 


\subsubsection{Flux Gate Magnetometers (MAG)}
\label{ssec:MAG}

\begin{figure}

    \centering

    \includegraphics[width=0.35\columnwidth]{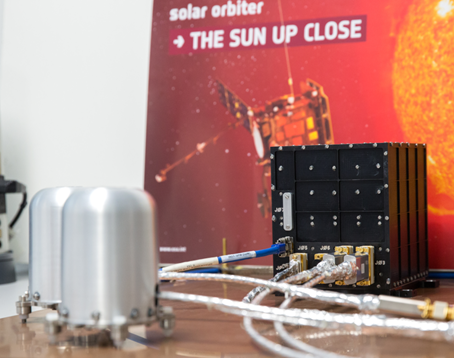}

    \caption{MAG is based on the flight-proven Solar Orbiter magnetometer design.}

    \label{fig:MAG}

\end{figure}

The MAG is a dual core fluxgate magnetometer designed and built by Imperial College London (Imperial) which will be carried on every HS SC to measure the local magnetic field. 
The MAG design is based on direct heritage from the successful Solar Orbiter \cite{Horbury:2020b} magnetometer (Fig.~\ref{fig:MAG}) with modifications taken from the soon-to-fly JUICE (JUpiter ICy moons Explorer) instrument. 
HelioSwarm MAG will carry just one sensor on each spacecraft, at the end of a dedicated 3~m boom to minimise the effects of spaceraft fields, connected to the instrument electronics box via a harness. 
The electronics box will contain a power supply and Front End Electronics board: the latter will drive the sensor and digitise the signal, sending it directly to the spacecraft digital processing unit (DPU) where it will be filtered and decimated to 16 vectors/s on a common timeline with the SCM.

MAG operations are straightforward, with the instrument operating throughout the science orbit. The instrument will have a 4~pT resolution in its most sensitive range of $\pm128$~nT but can range automatically up to $60,000$~nT and can therefore operate in a full Earth field before launch.

MAG data will be calibrated at Imperial College, with inter-calibration between S/C performed to ensure that derived products such as volumetric currents are reliably estimated.
MAG therefore contributes directly to the multi-point vector magnetic field measurement observable in Table~\ref{tab:observables}, but is also central to the AC magnetic field measurement as well as some plasma products such as temperature anisotropies.

\subsubsection{Search Coil Magnetometers (SCM)}
\label{sssec:SCM}


The SCM is a heritage set of magnetic sensors designed and built by Laboratoire de Physique des Plasmas (LPP) and Laboratoire de Physique et Chimie de l’Environnement et de l’Espace (LPC2E) selected to measure the IMF’s higher frequencies needed to capture advected ion-scale structures.
The HS SCM design is based on the most recent sensor developed for the ESA JUICE mission by LPP \cite{Wahlund:RPWI,Retino:SCM,Retino:2020}. 
LPP and LPC2E will be responsible for the testing and calibration of the instruments.

The SCM consists of a  tri-axial set of 20 cm long magnetic sensors with associated preamplifier (ASIC) mounted at the tip of a 3 m boom opposite to the MAG boom. 
Each sensor axis consists of two windings (a primary and a secondary) around an internal PEEK mandrel inside which the ferromagnetic core (mu-metal) used on other flight heritage missions, (e.g. Cluster \cite{Cornilleau-Wehrlin:2003} or THEMIS \cite{Roux:2008}) resides. 
Windings are connected to the preamplifier which drives the analog signal down the SCM boom harness to the IDPU which performs the digitization. 
SCM ground calibration is performed at the National Magnetic Observatory of Chambon-la-for\^et using a facility upgraded by LPP for MMS and BepiColombo.

Each primary winding response is modified through a flux-feedback applied via a secondary winding to produce the frequency response and phase stability needed for Observatory-level analyses. 


SCM has a single science operational mode drawing a steady 0.3 W. 
The three differential analog outputs of the SCM preamplifier are anti-alias filtered and digitized by the IDPU receiving electronics at 128 Sps then filtered to 32 Sps to satisfy HS observational requirements described in Table~\ref{tab:observables}.  
This science operational mode is only interrupted during the in-flight calibration sequence. 
This sequence, scheduled for one per orbit and following events such as maneuvers and eclipses, will follow procedures successfully implemented on MMS, PSP, and SolO. 
It is performed to assess the stability of the transfer function through the mission using a calibration signal provided by the IDPU.



\begin{figure}
    \centering
    \includegraphics[width=0.35\columnwidth]{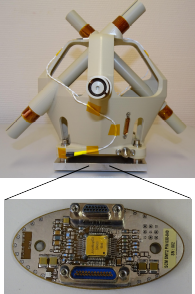}
    \caption{JUICE SCM heritage instrument with its ASIC preamplifier.}
    \label{fig:SCM}
\end{figure}

\subsection{Faraday Cups (FC)}
\label{ssec:FC}

The Faraday Cup (FC) is a heritage-based design developed at the Smithsonian Astrophysical Observatory (SAO), in conjunction with University of California, Berkeley (UCB), and Draper Laboratories.  The sensor makes measurements of the radial VDFs of SW ions along with the flow angle of the incoming beam to measure proton densities and velocities over the ranges and sensitivities typical of the pristine SW.

Previous Faraday Cups have been employed on a wide variety of missions including Voyager \cite{Bridge:1977}, Wind\cite{Ogilvie:1995}, DSCOVR\cite{Lotoaniu:2022}, and PSP\cite{Case:2020,Kasper:2015}.  Two of the HS FC electronics boards (the logic/signal analysis board and the low-voltage power supply) are direct copies of the PSP electronics.  A third board (the high-voltage power supply) is a fully-qualified backup design from the PSP instrument development.  The instrument uses an oscillating electric potential to create an electric field that accepts or rejects particles based on their energy/charge.  Particles with large enough E/q to successfully transit the electric field deposit their charge onto collector plates that measure the incoming current of SW particles.

A Faraday Cup instrument is placed on the sun-facing side of each spacecraft so that an unimpeded view is maintained in the direction of the Sun.  
Because Elsasser analysis involves measurements from both the IMF and SW, cross-sensor timing, pointing, and alignment requirements between the magnetometers and FC are levied.

The Faraday Cup operates in a single operational configuration throughout all phases of the mission.  The instrument starts at its lowest voltage (energy/charge) and steps its way upward through 16 voltage windows while making measurements of the incoming SW current on each of its four collector plates in each window.  The instrument keeps track of the maximum current measured in the previous spectrum so that the following spectrum can be measured with a more focused voltage range with better resolution.

The FC instrument design parameters have been determined by analyzing the historic distribution of all measurements made by the \textit{Wind} Faraday Cup instrument.  The aperture sizes, voltage ranges, and field-of-view for the HelioSwarm instrument are designed to capture more than 98\% of the SW conditions (velocities, densities, and temperatures) that have been previously observed.  The resulting voltage range will allow for measurement of proton velocities \change{in the range of} about 200-850 km/s.


The Faraday Cup instruments provide a measurement of the radial distribution functions of the SW plasma on each of the nine spacecraft along with plasma quantities derived from those distributions.  By calculating the moments of the distribution and by fitting an assumed functional form to the distribution, the vector velocity, density, and radial temperature can be provided.  These data products contain 8 measurements per second and fulfill the multi-point measurement requirements of the velocity and density of the SW, as shown in Table~\ref{tab:observables}.  


\begin{figure}
    \centering
    \includegraphics[width=0.5\columnwidth]{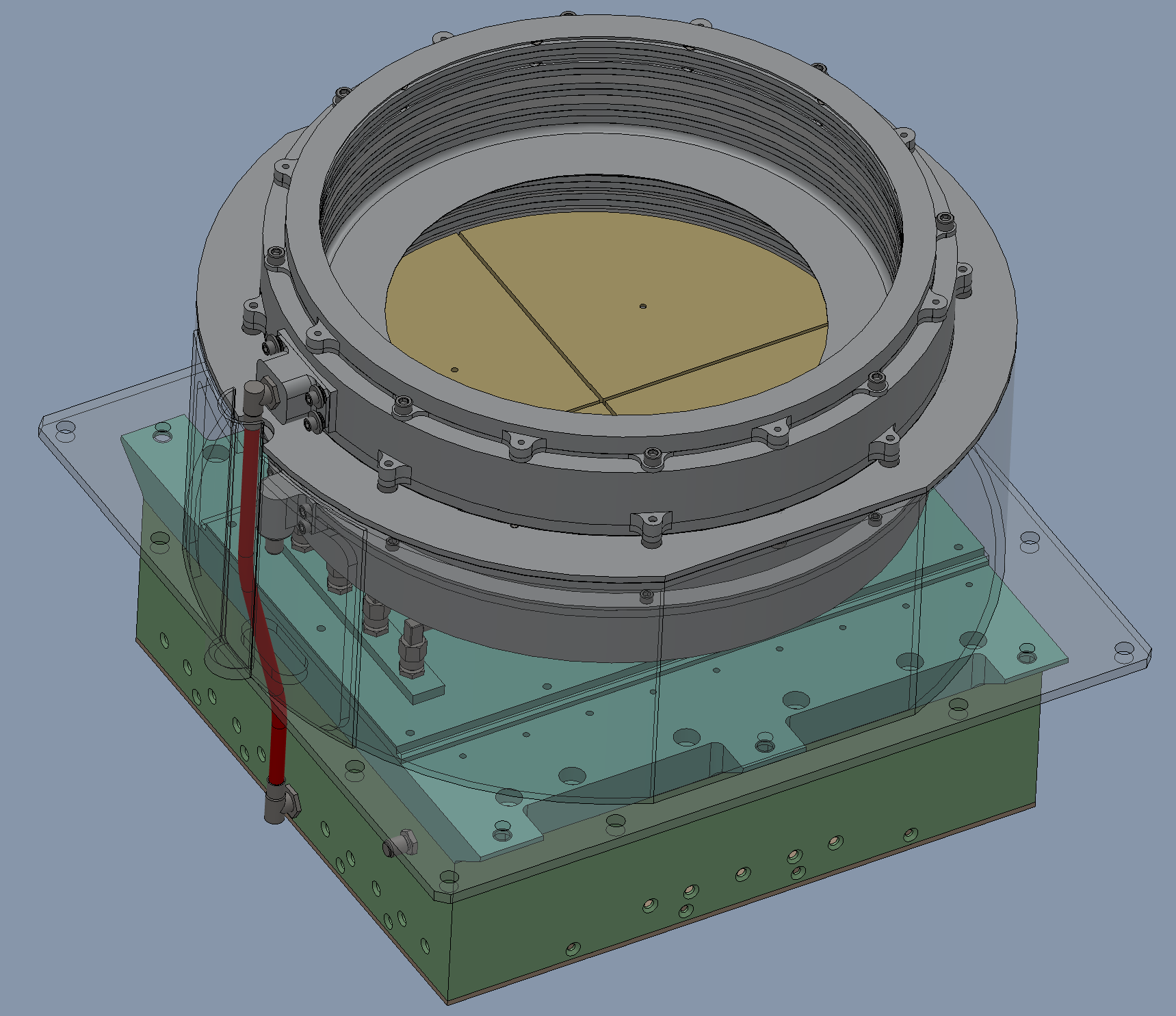}
    \caption{The Faraday Cup mechanical design (as of the concept study report).  The mounting bracket is displayed in a semi-transparent mode.  The instrument consists of two main subassemblies: the sensor (the top cylindrical section) and the electronics module (the lower rectangular box), which are connected by flexible coaxial cables.}
    \label{fig:FC}
\end{figure}

\subsection{Ion Electrostatic Analyzer (iESA)}
\label{ssec:iESA}


The iESA is a particle sensor designed and built by {Institut de Rescherch en Astrophysique et Planetologie} (IRAP, Toulouse, France), {Laboratoire d'Astrophysique de Bordeaux} (LAB, Pessac, France), University of New Hampshire (UNH, USA), and Mullard Space Science Laboratory (MSSL, UK), with IRAP technical leadership and heritage. 
The direct heritage instrument is the Proton and Alpha Sensor \cite{Owen:2020} onboard the Solar Orbiter mission \cite{Muller:2020}, with some sub-systems inherited from other past particle instrumentation led by IRAP (on STEREO, MAVEN, Cluster, etc.). 
iESA measures the full proton and alpha particle distribution functions with an unprecedented combination of high energy, angular and time resolutions (cf. Table~\ref{tab:observables}). 

\begin{figure}
    \centering
\includegraphics[width=0.35\columnwidth]{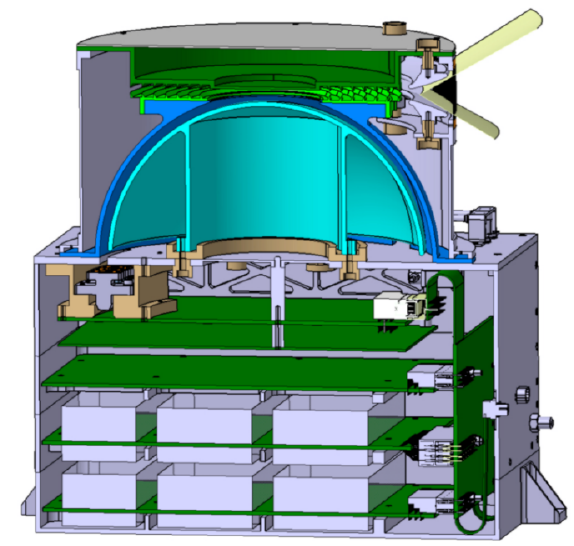}
    \caption{iESA subsystem components (as of the concept study report), include the deflectors, collimators, and analyzer spheres (top), as well as the 16 channel electron multipliers, front end electronics, low- and high-voltage power supply,  and FPGA board.}
    \label{fig:iESA}
\end{figure}

As illustrated in Figure~\ref{fig:iESA}, entrance deflectors allow for the sweeping of input elevation angles $\pm 24^\circ$ from the main detection plane with $3^\circ$ angular binning, which is resolvable with the use of a collimator. 
The deflected and collimated ions are then subject to E/q selection through a classic top-hat electro-static analyzer. 
The E/q selected ions are focused onto the main detection plane, which comprises 16 channel electron multipliers (CEMs). 
These perform a $10^7$ gain in charge collection on anodes with $3^\circ$ resolution in azimuth over an angular range of $\pm 24^\circ$ as well, allowing for a homogeneous $\pm 24^\circ$ field-of view with $3^\circ$ angular resolutions, in both elevation and azimuth angles. 
The iESA electronics contains (1) a front-end board comprising 16 CEMs with associated anodes and amplificators, (2) a high-voltage board to supply the entrance deflectors, analyzer plates, and CEMs with the required (static or sweeping) high voltages, as well as (3) an FPGA and (4) a low-voltage power supply board dedicated to instrument control and power.


iESA operations are based on the sequential stepping of the electrostatic analyzer and entrance deflector high voltages. 
The instrument implements SW beam-tracking strategies \cite{Owen:2020}, using previous measurements, to dynamically set the energy and angular bins of the next sample, allowing for faster measurement cadence. 
The iESA is highly versatile and the tracking strategy can implement any combination of energies and angles. 
Instrument operation will be adapted to the science target, but a primary operation mode is expected to be a Proton Tracking mode measuring the 3D VDFs of SW protons with high energy (8\%) and angular ($3^\circ$) resolutions at a cadence down to 150 ms, well into the sub-ion timescales. 
To characterize alphas, a Proton-Alpha Tracking mode will be used, with 48 energy bins and a 450 ms cadence, though longer accumulation times can be used to enhance counting statistics as needed.



\subsection{Student Electron Electrostatic Spectrograph Student Collaboration}
\label{ssec:eESA}

In addition to the magnetic field and ion instruments previously listed in this section, HS has also proposed to include a
Student Electron Electrostatic Spectrograph (SEE) Student Collaboration project to measure ambient, low energy electrons.
This electron instrument would be mounted on the Hub SC, and would be used to study the connectivity of the local magnetosphere, solar wind, and cis-lunar space via measurements of low-energy electron populations.
The project would be co-led by graduate and undergraduate students, with the prime deliverable from the SEE project a cohort of future scientists educated in the lifecycle of a NASA mission, including instrument development and merger of science goals with hardware design.
A backup design for SEE has PSP and ESCAPADE flight heritage \cite{Whittlesey:2020}.

\subsection{The Hub}
\label{ssec:Hub}

\begin{figure}
    \centering
    \includegraphics[width=0.85\columnwidth]{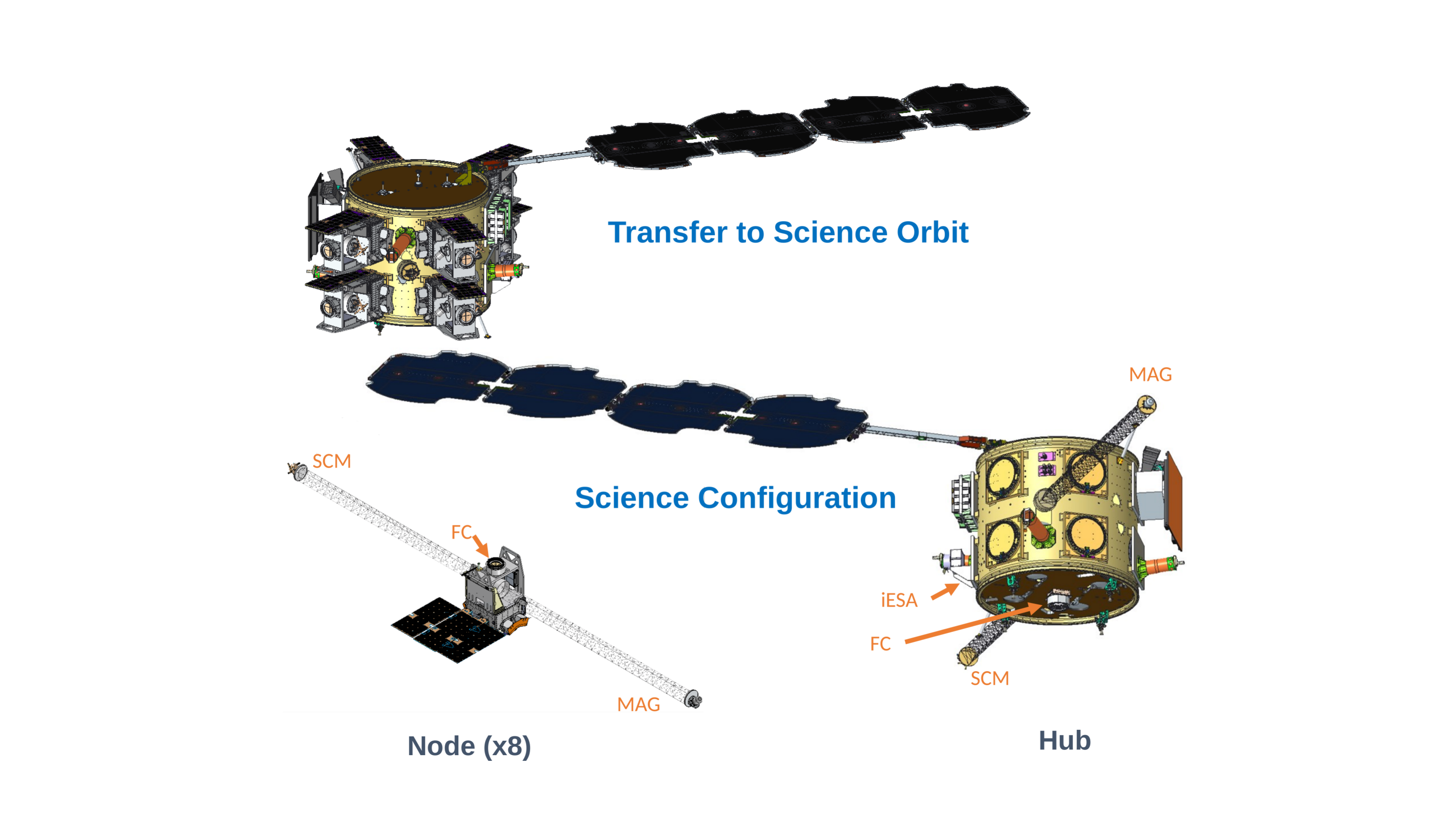}  
    \caption{Transfer (top) and Science (bottom) Configurations for the Hub (Sec.~\ref{ssec:Hub}) and Node (Sec.~\ref{ssec:Nodes}) S/C.}
    \label{fig:Hub}
\end{figure}

Northrop Grumman (NG) provides the Hub spacecraft. 
This ESPA-class spacecraft serves as the central relay for all Nodes within the Observatory and is based on the high-heritage ESPAstar line which was designed to carry separable payloads to orbit. 
The Hub is 730 kg at launch, including the hydrazine propellant necessary for carrying it and all the Nodes into the HS science orbit, illustrated at the top of Fig.~\ref{fig:Hub}.
The Hub is capable of generating 1165 W of power via its single deployable solar array. 
As configured for science, the Hub spans a maximum dimension of 8.4 m.

\subsection{The Nodes}
\label{ssec:Nodes}


The Node spacecraft are Blue Canyon Technologies (BCT) Venus-class spacecraft with standard accommodations for hosting the HS payload. 
As configured for HS, the SmallSat Nodes are just over 70 kg each and use onboard propulsion to maintain the proper swarm geometry. 
The Nodes generate 200 W of power and provide a single mechanical interface to the HS payload. 
With booms deployed, bottom left of Fig.~\ref{fig:Hub}, the maximum tip-to-tip dimension of the Nodes is just over 6m.



\subsection{Observatory Architecture}
\label{ssec:system}

The HS flight system will launch with the Hub carrying all 8 Nodes through a series of phasing loops and a lunar swingby into the science orbit over a period of approximately 72 days. 
Once the science orbit (a High Earth Orbit P/2 lunar resonant orbit with apogee greater than $60 R_E$ and perigee less than $12 R_E$), pairs of Nodes are separated from the Hub spacecraft and instrument checkouts and calibrations occur during the $\approx 81$-day commissioning phase.  
Once the Nodes have been maneuvered into their proper locations in the observatory, the 12-month science phase begins. 
HS uses a Hub-and-spoke communications architecture 
in which the Hub is the only direct link to the ground (via S-band, DSN) and each Node receives commands from and relays science data directly to the Hub via S-band crosslinks. 

The Mission Operations Center (MOC) is hosted out of ARC, with engineering support centers for Hub and Nodes at NG and BCT respectively. 
The Science Operations Center (SOC) is hosted out of the PI-institution, UNH, which is also responsible for delivering L1-L4 data products to NASA SPDF; see Sec.~\ref{ssec:data}.  
The HS missions operations approach strategically uses the inherent orbital dynamics of the observatory; the swarm naturally “expands” and “contracts” over each 14-day orbital period. 
The Flight Dynamics team has designed the staggering of the Node-Hub closest approaches around perigee to facilitate periods of high-rate data downlink for each Node each orbit. 
As the observatory begins to expand out towards apogee, the polyhedral performance, Sec.~\ref{sssec:poly}, requirements are satisfied. At apogee and on the contraction back in towards perigee, the longest-baseline 3D performance requirements are satisfied, Sec.~\ref{sssec:3D}.


\subsection{Data Processing and Selection}
\label{ssec:data}

The Science Data Center (SDC) at the UNH SOC supports HS data processing, produces timing and other ancillary data that are provided to the instrument teams, releases and archives L0-L4 data with associated calibrations, and provides data selection tools to the community.

Automated processes transfer L0 telemetry data from the HS MOC to the SOC within 24 hours of ground receipt.
Upon L0 receipt, the SOC performs packet format checks (e.g., valid headers, checksums).
The SOC prepares a timing product to correct Node clock differences relative to the Hub timing and processes IS housekeeping (HK) to calibrated units (L1). 
L0 data are provided to the instrument teams for processing, with timing corrections and L1 HK included as additional inputs.
The instrument teams generate and validate L1 (measurements in engineering units), L2 (science data---magnetic field measurements, particle velocity distributions and the associated moments and fits--- in payload co\"ordinates), and L3 (science data in RTN co\"ordinates) data products within 30 days of receipt.

L1-L3 data are retrieved by the SOC from the instrument teams within 24 hours of processing and summary plots generated. 
L4 data products, e.g. a merger of the MAG and SCM data or combined magnetic field and proton products from across the observatory, are produced at the SOC within 5 working days.

Upon completion of IS commissioning, the SOC begins a period of data product and instrument performance validation. 
As data for each orbit are downlinked, they are routinely and automatically processed. 
During the subsequent orbit, the SDC lead co\"ordinates instrument team validation of the data. 
Validation activities proceed through individual calibration and Observatory-level calibrations. 
As calibrations are updated, previous orbits' data are reprocessed with the updated calibrations, so the process iterates with increasing data volumes of increasing refinement.
Validated, calibrated L2 through L4 data are provided to NASA SPDF for public access upon completion of the validation period, anticipated to be no more than 5 months, which may be shortened if reasonable calibrations are available sooner.

Because HS comprises nine spacecraft with slowly changing relative positions in an equally dynamic plasma environment, specialized tools to assist researchers in their data selection will be developed \change{and implemented} at the SOC. 
Many techniques employed by the researchers require specific Observatory configurations. 
To aid the researcher in the selection of processed data, the SOC is developing interactive queries, but {all science data are downlinked and processed regardless of observatory configuration}.
As an example, Figure~\ref{fig:example} shows a snapshot from a preliminary data selection code using DRM orbit trajectories. 

\section{Analysis Approaches}
\label{sec:analysis.methods}

A variety of multi-spacecraft Analysis Approaches have been previously developed for missions such as Cluster and MMS.
These methods include calculations of
\begin{itemize}
\item \textbf{Cascade Rate}, measuring of the transfer of turbulent fluctuation energy from one spatial scale to another \cite[e.g.][]{Hadid:2018,Pecora:2023},
\item \textbf{2-point correlation}, measuring the temporal and spatial scale over which a spectral element is remade by nonlinear processes \cite[e.g.][]{Horbury:2000,Matthaeus:2016},
\item \textbf{Structure Functions}, determining the statistics turbulent field increments to reveal scale-dependent, intermittent turbulence\cite[e.g.][]{Sreenivasan:1997,Chhiber:2021},
\item \textbf{Wave telescope}, determining the wavevectors of
plasma waves and their associated 3D power distributions\cite[e.g.][]{Pincon:1991,Narita:2022},
\item \textbf{Pressure-strain interaction}, measuring the dilation, $-(\underline{\underline{P}}\cdot \nabla )\cdot \V{U}$, which describes the local conversion between flow and thermal energy\cite[e.g.][]{Yang:2017,Cassak:2022},
\item \textbf{Curlometer \& Gradient Methods}, which construct current sheets and other intermittent structures from spatially distributed
measurements\cite[e.g.][]{Dunlop:2002b,Dunlop:2021}.
\end{itemize}
Many of these methods and their applications have been documented in ISSI review articles over the last several decades\cite{Paschmann:1998,Paschmann:2008}.
In this section, we apply some  analysis approaches to synthetic timeseries constructed using DRM trajectories corresponding to selected HS configurations through different numerical simulations of turbulence including two-fluid \cite{Wang:2015,Juno:2018} and hybrid-PIC \cite{Kunz:2014,Arzamasskiy:2019} nonlinear simulations and quasilinear simulations \cite{Klein:2012,Klein:2014a}.
We note that due to computational limitations, all the numerical codes used make approximations in terms of the physical processes included and/or the scales simulated. 
Therefore, we do not expect the numerical simulations to be completely representative of actual plasma turbulence at all scales, and differences between simulations and HS observations will drive improvements of the modeling of turbulent transport and dissipation.

\subsection{Multipoint Correlations and Structure Functions}
\label{ssec:analysis.msa}

Multipoint spectral analysis, 2nd-order structure functions, and space-time correlation functions yield distributions of turbulent energy in configuration space \cite{Chen:2011a,Paschmann:1998,Paschmann:2008} and, through time-lagging, decorrelation times for fluctuations at measured spatial scales \cite{Matthaeus:2016}.

\begin{figure}
    \centering
    \includegraphics[width=0.75\columnwidth]
    {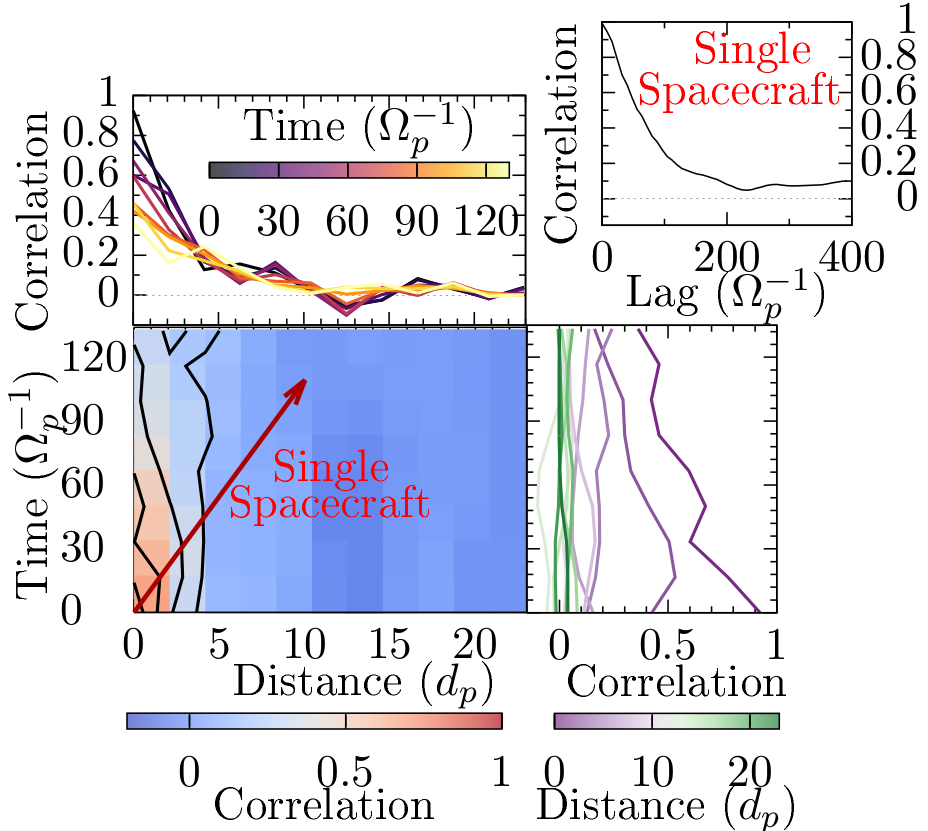}
    \caption{
    The spatial-temporal correlation calculated from synthetic HS measurements extracted from a hybrid-PIC simulation of turbulence\cite{Arzamasskiy:2019}. By using all nine trajectories, we are able to resolve the spatial and temporal dependences of Eqn.~\ref{eqn:R.def} independently, unlike the auto-correlation from a single trajectory, presented in the upper right-hand panel, which effectively samples along the red arrow in the lower left panel, convolving together spatial and temporal variations.}
    \label{fig:RT_corr}
\end{figure}

Figure~\ref{fig:RT_corr} illustrates the temporal and spatial decorrelation of signals from synthetic magnetic field data set constructed by sampling a hybrid-PIC numerical simulation \cite{Arzamasskiy:2019} over trajectories defined by the DRM described in Section~\ref{ssec:orbits}.
The correlation is calculated between all nine trajectories at each point in the timeseries as
\begin{equation}
  R(\vert\V{r}\vert,\tau)=
  \left<\V{b}(\V{x},t)\cdot\V{b}(\V{x}+\V{r},t+\tau) \right>.
  \label{eqn:R.def}
\end{equation}
These correlation values, and how quickly they depart from unity, characterizes the temporal and spatial scales over which fluctuations are remade by nonlinear terms, and represents a key statistical property of turbulent distributions \cite{Matthaeus:2005} that will be used in G1O1 and G1O2.
For comparison, an example auto-correlation produced using lagged timeseries from a single trajectory (upper right panel) mixes together spatial and temporal dependence. 
HS will disentangle spatial and temporal correlations which single SC convolve together.

The measurement of intense fluctuations at smaller scales while simultaneously measuring the distribution of fluctuations at larger scales differentiates between models of scale-dependent intermittency, testing theoretical predictions\cite{Tennekes:1975,Grauer:1994,Chandran:2015,Mallet:2017a,Greco:2009}.
Intermittency also affects different types of proton heating mechanisms in different ways, with the associated coherent structures greatly affect the efficiency of certain processes. 
For example, stochastic heating occurs when fluctuation amplitudes at the scale of a particle's gyroradius become large \cite{Chandran:2010a,Dalena:2014}, which are enhanced near coherent structures.
Conversely, dissipation mechanisms such as Landau damping are less affected by intermittency \cite{Mallet:2019}.
HS multiscale measurement of higher order intermittent statistical measures, along with temperature and temperature anisotropy reveal deep connections between cascade, intermittency and dissipation.


\begin{figure}
    \centering
    \includegraphics[width=\columnwidth]{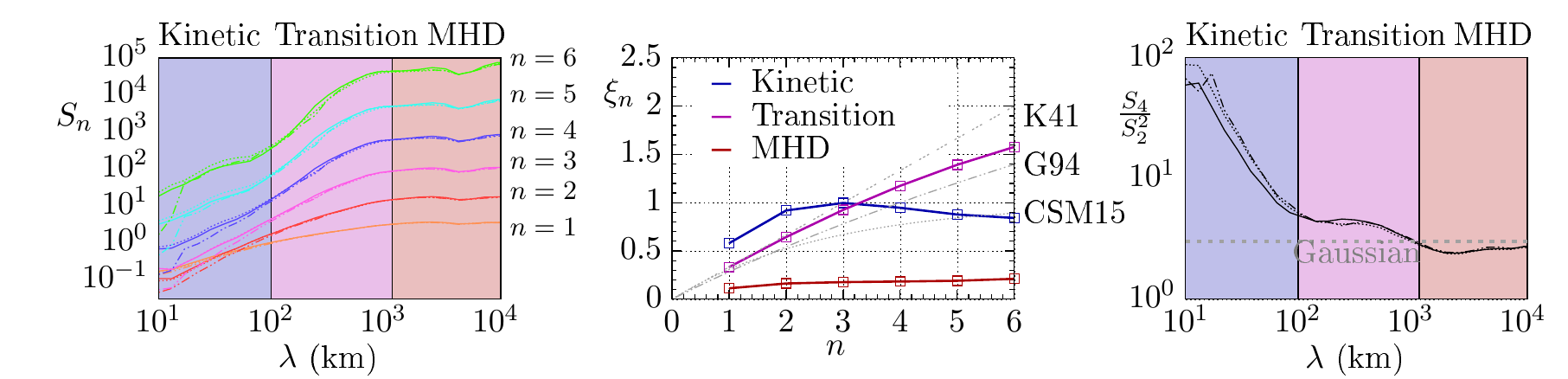}
    \caption{HS's multiscale configuration enables the calculation of multiple orders of structure functions (left panel), that when fit to a power-law $S_n \propto \lambda^{\xi_n}$ (center) or recombined into quantities such as kurtosis $S_4/S_2^2$ (right) can be compared to theoretical predictions [e.g. \cite{Kolmogorov:1941,Grauer:1994,Chandran:2015}] to characterize the scale-dependent intermittency in turbulence (G1O4).}
    \label{fig:intermit_2}
\end{figure}

\begin{figure}
    \centering
    \includegraphics[width=0.5\columnwidth]
    {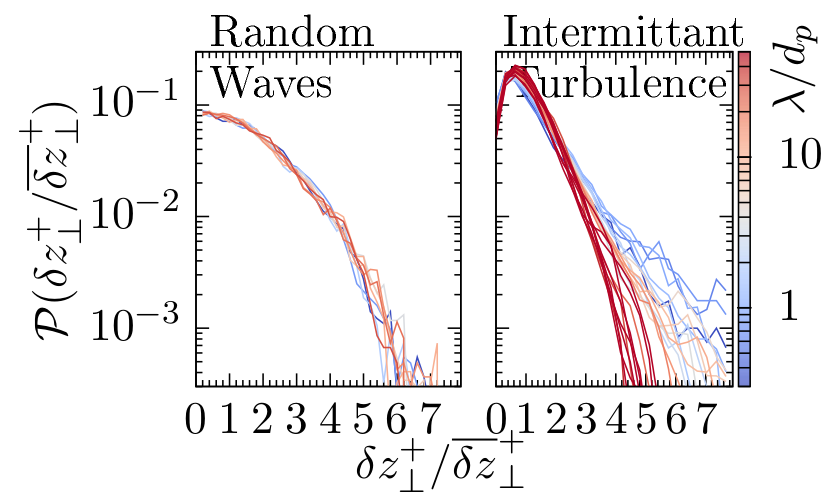}
    \caption{HS measures intermittency at MHD (red) and
sub-ion (blue) scales simultaneously from different numerical simulations, providing science closure on G1O4 by differentiating between models of nonlinear coupling that predict enhanced amplitudes of normalized Elsasser fluctuations $\delta z^+_\perp$ at small scales (right, from a turbulent hybrid-PIC simulation \cite{Arzamasskiy:2019}) compared to a scale-independent normal distribution of amplitudes (left, drawn from ensemble of randomly-phased wave modes.).}
    \label{fig:intermit_dz}
\end{figure}

One means of quantifying the intermittency of a system is illustrated in Figure~\ref{fig:intermit_2}, where we calculate the scale-dependent structure function $S_n$ using synthetic times series drawn from the same hybrid numerical simulation of turbulence \cite{Arzamasskiy:2019}. 
Instead of using increments drawn from a single timeseries
\begin{equation}
  S_n (\lambda=\tau v_{{SW}} ) =\left<[\delta z(t+\tau)-\delta z(t)]^n \right>
\end{equation}
\cite{Chhiber:2021}, increments are calculated using all nine timeseries combined with HS’s spatial resolution and configurations and a modified form of Taylor's hypothesis \cite{Horbury:2000} allowing for orders of magnitude more samples to be used at a given scale both along and transverse to
 the magnetic field and flow directions, 
\begin{equation}
  S_n [\V{x}_i(t+\tau) - \V{x}_j(t)]= \left<\left\{\delta z[\V{x}_i(t+\tau)]-\delta z[\V{x}_j(t)]\right\}^n \right>
\end{equation}
 enabling the
 calculation of the higher order structure functions spanning ion kinetic (blue, left panel), transition (purple), and MHD (red) scales; for N increments measured, structure functions of order $n=\log_{10} [N]-1$ can be resolved \cite{Dudok:2004,Dudok:2013}; HS enables the study of higher order $S_n$ than previous missions, where differences between theoretical descriptions are more easily distinguishable.
 From these measurements of $S_n$, theories about intermittency as a function of scale \cite{Sreenivasan:1997}, which describe how frequently and how abruptly sharp structures of different sizes and shapes arise, are tested by fitting $S_n \propto \lambda^{\xi_n}$ over different scale ranges (center); the trending of the fit parameters with order n is
used to validate, falsify, or improve theories [e.g. \cite{Kolmogorov:1941,Grauer:1994,Chandran:2015}. 
Combined with other metrics such as kurtosis (right panel) as well as the analysis method presented in Figure~\ref{fig:intermit_dz}, HS characterizes turbulent intermittency as a function of scale addressing G1O4.

Fig.~\ref{fig:intermit_dz} illustrates with the same simulated synthetic DRM timeseries used for Fig.~\ref{fig:RT_corr} another way in which HS will provide science closure on G1O4 by using its 36 baseline separations to simultaneously quantify the distribution of turbulent fluctuations at large and small scales.
Following Mallet et al 2015\cite{Mallet:2015}, we define the fluctuation amplitude increment
\begin{equation}
\delta z_{\perp}^{\pm}=
\vert \delta \V{z}_{\perp}^{\pm} \vert=
\vert \V{z}_{\perp}^{\pm}(\V{r}_0+\V{r}_{\perp})-\V{z}_{\perp}^{\pm}(\V{r}_0)\vert
\end{equation}
where $\V{r}_\perp$ is the separation in the plane perpendicular to the mean magnetic field direction $\V{B}_0$. 
Using both the synthetic HS timeseries drawn from a hybrid-PIC simulation, as well as timeseries constructed from a collection of randomly phased waves, we calculate
$\delta z_{\perp}^{\pm}$ as a function of scale $\lambda = \vert\V{r}_\perp\vert$ for an ensemble of separations.
We then calculate the median value of the increment $\delta \overline{z}_{\perp}^{\pm}$ over a series of bins spaced logarithmicaly in $\lambda$, and normalize the probability distribution function of the increments in each bin by the median.
For the trivial case of randomly-phased waves, the increments are normally distributed, and there is no variation as a function of scale.
For the case of simulated turbulent, the intermittency increases with decreasing scale, seen in the transition from red (large, MHD scales) to blue (smaller, ion-kinetic scales).
As the intermittency of a turbulent system depends on the nature of the nonlinear interactions that transport energy through a system, these types of distributions represent a sensitive test of different models of turbulence.

\subsection{Curlometer and Gradient Techniques}
\label{ssec:analysis.wave}

    Wave telescope and curlometer techniques reveal the nature of fluctuations and identify structures within turbulence. 
    Gradient estimation is closely related and is required for determining the pressure-strain tensor and the production rate of internal energy. 
    

   


   
   \begin{figure}
    \centering
    \includegraphics[width=\columnwidth]{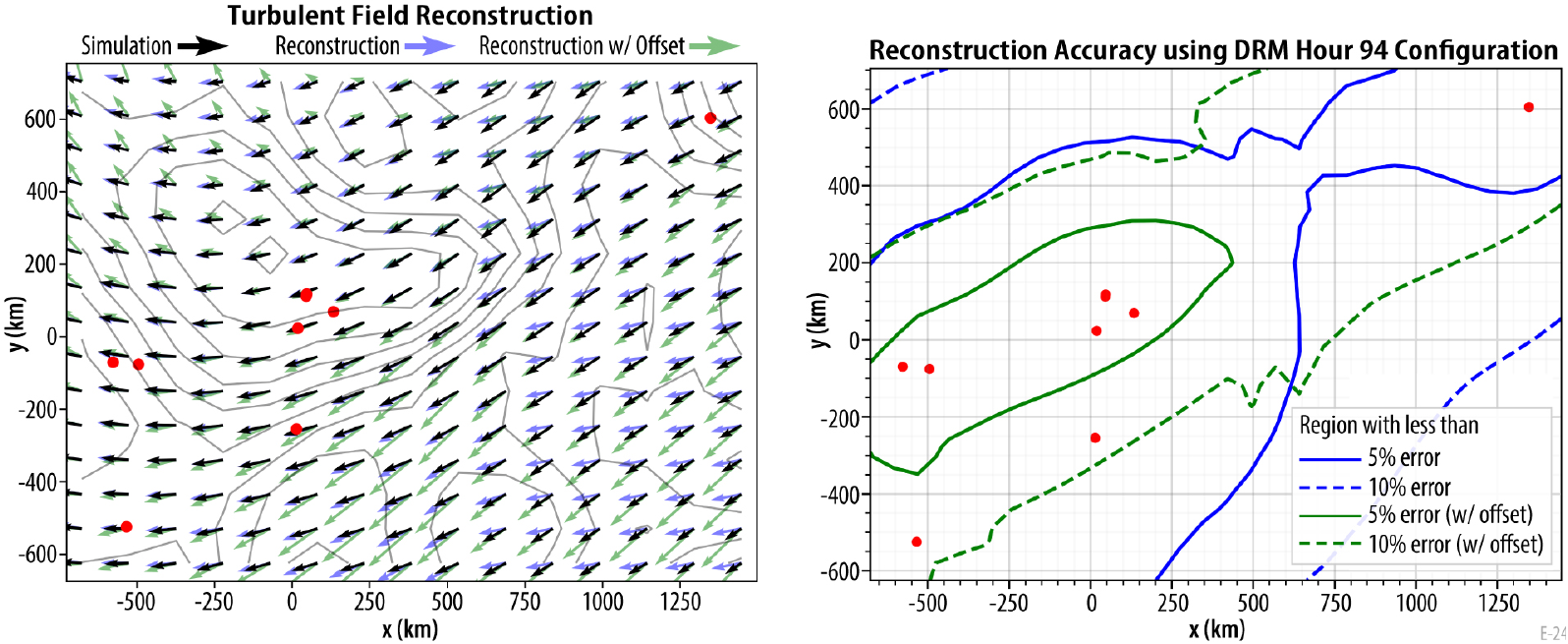}
    \caption{HS's SC positions (red points) enable a high-fidelity reconstruction (perfect measurements shown in blue,
 measurements with included systematic errors in green) of the magnetic field (arrows) over a larger volume of space, enabling the simultaneous
study of ion-scale structures (e.g., current sheets, shown as contours at left) with sufficient accuracy to address G1O4 (regions of
reconstruction with less than 5\% and 10\% error shown at right, \change{as explained in the text}).}
    \label{fig:reconstruction}
\end{figure}

As an example of a novel application of gradient techniques enabled by a multi-spacecraft observatory, estimates for the spatial gradients from synthetic timeseries drawn from a two-fluid numerical simulation of turbulence\cite{Wang:2015} are used in a first-order reconstruction to reconstruct a 3D synthetic magnetic field, shown in Figure~\ref{fig:reconstruction} and described in detail in \cite{Broeren:2021}. 
By using a weighted average of the first-order estimates for the reconstructed magnetic field drawn from the spatial gradients determined from the 126 unique tetrahedra that comprise the HS Observatory, we accurately reproduce the magnetic field over large volumes of the simulation. 
For a selected DRM interval in a good polyhedral configuration, we reconstruct the simulated turbulent magnetic field (black/blue arrows indicate the simulated/reconstructed field). 
Averaging over many spatial locations in the simulation, this analysis method yields $\leq 5$\% relative error over a volume of nearly $1.82\times 10^9 \textrm{km}^3$ (solid blue line, right panel) and $\leq 10$\% error over a volume of $3.208\times 10^9$ km$^3$ (dashed blue line). 

To quantify the impact of systematic errors on this analysis technique, we introduce random offsets replicating estimated systematic errors to each component of the measured magnetic field at all nine SC (green lines). 
 This produces reconstructed volumes of $3.66\times 10^8 (1.77\times10^9)$ km$^3$ for the 5\% (10\%) error thresholds.
 By leveraging the large number of tetrahedral configurations sensitive to many length scales simultaneously, HS enables simultaneous studies of both MHD-scale structure as well as much smaller current sheets, bringing closure to questions about the transfer of energy from fields to flows (G1O3) and associated heating near ion-scale intermittent structures (G1O4).

As discussed at the beginning of this section several other multispacecraft analysis methods, previously implemented on missions such as MMS and Cluster, can be immediately applied to HS observations, or extended to incorporate information from all nine spacecraft in the observatory, e.g. \cite{Perri:2017,Zhang:2021,Pecora:2023,Toepfer:2023,Broeren:2023}.

\section{Conclusions}
\label{sec:conclusions}

Turbulence is the process by which energy contained in fluctuating magnetic fields and plasma motion cascades from larger to smaller spatial scales, and ultimately into thermal energy of charged particles comprising the plasma. 
In addition to being a key process that heats cosmic plasmas, it also creates the conditions in which all universal plasma processes (e.g., magnetic reconnection, shocks, particle acceleration) act, both within the heliosphere and in all astrophysical domains. 
Due to its fundamentally multiscale nature, only spatially distributed, simultaneous measurements provide the data needed to bring closure to outstanding questions about the distribution and transfer of turbulent energy.
HS achieves its mission objectives through an innovative swarm implementation of high-heritage mission elements, ranging from instruments, to spacecraft, operations, and analysis tools.
HS provides a paradigm shift in mission design where the many elements of the swarm and the way they interact form an Observatory that is far more than the sum of its parts.
As the first multipoint, multiscale mission, HS gives an unprecedented view into the nature of space plasma turbulence.

\textbf{Acknowledgements} We are deeply indebted to the incredible members of the HelioSwarm science, engineering, and proposal teams whose tireless efforts enabled this mission.
Construction and analysis of the HelioSwarm Observatory Design Reference Mission trajectories was supported in part by the HelioSwarm Project funded under NASA’s Prime contract no. 80ARC021C0001.






\end{document}